\RequirePackage{xr-hyper}
\documentclass[referee,sn-nature,oneside]{sn-jnl}

\usepackage{graphicx}%
\usepackage{multirow}%
\usepackage{amsmath,amssymb,amsfonts}%
\usepackage{amsthm}%
\usepackage{mathtools}
\usepackage{mathrsfs}%
\usepackage[title]{appendix}%
\usepackage{xcolor}%
\usepackage{textcomp}%
\usepackage{manyfoot}%
\usepackage{booktabs}%
\usepackage{algorithm}%
\usepackage{algorithmicx}%
\usepackage{algpseudocode}%
\usepackage{listings}%
\usepackage{lineno}
\usepackage{subcaption}
\usepackage{placeins}
\theoremstyle{thmstyleone}%
\newtheorem{theorem}{Theorem}

\theoremstyle{thmstyletwo}%

\theoremstyle{thmstylethree}%

\begin{document}

\title[COATI multiomics TI]{Synchronized Optimal Transport Reconstructs Coupled Transcriptomic and Chromatin Dynamics}

\author[1]{\fnm{Jingfeng} \sur{Wang}}
\author*[2]{\fnm{Yanxiang} \sur{Zhao}}\email{yxzhao@email.gwu.edu}
\author*[3,4]{\fnm{Qing} \sur{Nie}}\email{qnie@uci.edu}
\author*[1,5]{\fnm{Zixuan} \sur{Cang}}\email{zcang@ncsu.edu}

\affil[1]{\orgdiv{Department of Mathematics}, \orgname{North Carolina State University}, \orgaddress{\city{Raleigh}, \state{NC}, \country{USA}}}
\affil[2]{\orgdiv{Department of Mathematics}, \orgname{George Washington University}, \orgaddress{\state{Washington D.C}, \country{USA}}}
\affil[3]{\orgdiv{Department of Mathematics}, \orgname{University of California, Irvine}, \orgaddress{\city{Irvine}, \state{CA}, \country{USA}}}
\affil[4]{\orgdiv{Department of Systems Biology}, \orgname{University of California, Irvine}, \orgaddress{\city{Irvine}, \state{CA}, \country{USA}}}
\affil[5]{\orgdiv{Center for Research in Scientific Computation}, \orgname{North Carolina State University}, \orgaddress{\city{Raleigh}, \state{NC}, \country{USA}}}

\abstract{
Single-cell RNA and ATAC data collected across developmental stages provide time-course snapshots of transcriptional and regulatory states. Since sequencing is destructive, trajectories between these snapshots must be inferred from independently sampled cell populations. Existing approaches typically infer continuous dynamics in a single modality or integrate RNA and ATAC in a shared representation. Here we present Cross-Omics Aligned Trajectory Inference (COATI), which extends unbalanced synchronized dynamical optimal transport to time-dependent cross-modal maps for reconstructing coupled trajectories in distinct RNA and ATAC state spaces.
We evaluated COATI on synthetic benchmarks and four developmental datasets using held-out-stage reconstruction and lineage analyses. COATI improved reconstruction of cell populations at held-out stages, including recovery of chromatin-state heterogeneity within cell types, and better preserved developmental lineage relationships. The unbalanced formulation allows the inference of population mass changes, which were shown to associate with DNA-replication programs at early stages and lineage-maturation programs later in development.
The coupled trajectories enabled analysis of lineage-specific transcriptional and chromatin changes and their temporal relationships during development. Lineage responses inferred by COATI agreed with independent CRISPR perturbation results, and the inferred regulatory patterns were consistent with independent epigenetic data.
}


\maketitle

\section*{Introduction}
Gene expression and chromatin accessibility both change as cells differentiate. Single-cell RNA sequencing (scRNA-seq)~\cite{jovic2022single} measures expression and single-cell ATAC sequencing (scATAC-seq) measures accessibility~\cite{minnoye2021chromatin}. Multiome assays capture both in the same cells~\cite{cao2018joint}. Together, joint analysis of the two modalities makes it possible to study how changes in one modality are associated with changes in the other.

Because single-cell measurements are destructive, each cell is measured only once, and how individual cells change over time has to be inferred~\cite{weinreb2018fundamental}. Pseudotime trajectory inference methods use similarities in gene expression to reconstruct developmental paths and order cells along them by assigning a pseudotime to each cell~\cite{trapnell2014dynamics,wolf2019paga,saelens2019comparison}. Further using information of unspliced and spliced RNA abundances, RNA velocity estimates local directions of gene expression changes~\cite{la2018rna}. Time-course data provide a sequence of cell state distributions at known measurement times on separate samples. Waddington-OT estimates probabilistic couplings between distributions at successive time points using static optimal transport, which links earlier cell states to their possible descendants~\cite{schiebinger2019optimal}.

To model population dynamics continuously between sampled stages, dynamical optimal transport (OT)~\cite{benamou2000computational} was used to represent changes in cell state distributions as a continuous flow, which reconstructs a continuous trajectory for each cell between snapshots. TrajectoryNet~\cite{tong2020trajectorynet} uses neural ordinary differential equations~\cite{chen2018neural} to solve a regularized dynamical OT problem. Other dynamical OT methods have been developed to keep trajectories close to data manifold structure, allow population mass to change, and include stochastic dynamics~\cite{huguet2022manifold,sha2024reconstructing,zhang2024learning}. Recent works have also used flow matching as a simulation-free efficient solver~\cite{tong2023improving, wang2025joint,peng2026wfr,peng2026multiscale,kapusniak2024metric}. However, these methods typically infer trajectories in a single molecular state space, usually RNA, without incorporating information from other important modalities, such as ATAC. Cell states that are close in RNA space may be far apart in ATAC space, and thus a trajectory that is plausible in RNA space may be implausible in ATAC space. Moreover, these methods do not reconstruct corresponding trajectories in ATAC space to describe how chromatin accessibility evolves alongside gene expression.

Most methods that jointly analyze RNA and ATAC data do not model continuous dynamics of cell changes. Multimodal integration methods such as MultiVI, scMVP and scDART align RNA and ATAC data in a shared latent representation~\cite{ashuach2023multivi,li2022scmvp,zhang2022scdart}, and scPair learns mappings between separate RNA and ATAC representations~\cite{hu2024scpair}. MultiVelo and related models describe gene-level kinetics linking chromatin accessibility and transcription~\cite{li2023multivelo,li2025multivelovae,hong2026moflow,avdeeva2026archvelo}, GraphVelo extends velocity estimates across molecular representations~\cite{chen2025graphvelo}, and DynaVelo learns the dynamics of gene expression and transcription factor motif accessibility using neural ODEs~\cite{karbalayghareh2025dynavelo}. 
For time course data, scTIE jointly learns a shared RNA and ATAC representation and estimates OT couplings between cells at consecutive time points~\cite{lin2024sctie}, and scMultiNODE learns continuous population dynamics in a shared RNA and ATAC latent space~\cite{2024scmultinode}.
Our approach instead models synchronized transport in separate RNA and ATAC spaces, allowing the geometry of both modalities to explicitly shape the inferred trajectories.

Building on our recent work on synchronized optimal transport (SyncOT) and its unbalanced extension (UnSyncOT)~\cite{cang2025synchronized,cang2026unsyncot}, we develop Cross-Omics Aligned Trajectory Inference (COATI) to reconstruct coupled population dynamics in RNA and ATAC spaces. COATI extends the formulation to time-dependent cross-modal maps and develops a neural network implementation for high-dimensional single-cell embeddings. In COATI, observations in both spaces directly constrain trajectory inference, allowing chromatin information to shape the inferred dynamics and enabling joint downstream interpretation of RNA and ATAC states. We evaluated COATI on synthetic benchmarks and four developmental datasets including mouse gastrulation, pancreatic endocrinogenesis, palate development and human cerebral organoid development. Compared with existing methods, COATI more accurately reconstructs cell state distributions at held-out stages, better captures chromatin heterogeneity within cell types, and more faithfully preserves developmental lineage relationships. In the gastrulation analysis, predictions from CRISPR-perturbed progenitors reproduce the observed preservation of the neural branch. In addition to trajectory reconstruction, mass changes inferred by COATI show stage-dependent associations with cell-cycle and differentiation programs. The paired RNA and ATAC trajectories also suggest temporal relationships between gene expression and chromatin accessibility at candidate peak-gene pairs along specific developmental lineages. Independent epigenetic data further support the biological relevance of candidate regulatory regions examined in these analyses.

\section*{Results}

\subsection*{Overview of COATI}
\label{sec:coati-overview-section}

\begin{figure}[!htbp]
    \centering
    \includegraphics[width=0.7\textwidth,keepaspectratio]{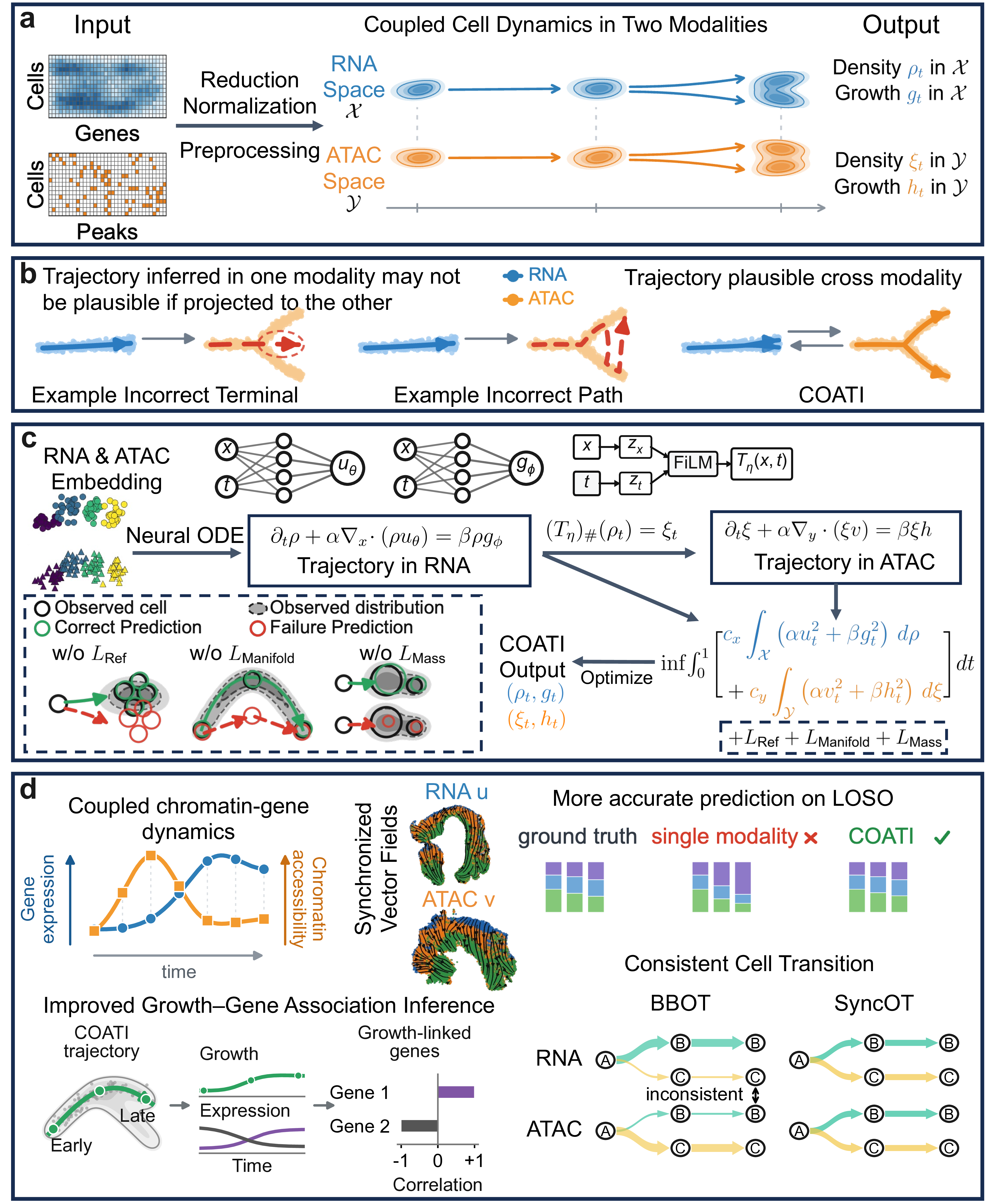}
    \caption{\textbf{Overview of COATI.}
    \textbf{a,} Cell-by-gene and cell-by-peak count matrices are processed and projected into reduced-dimensional RNA ($\mathcal X$, blue) and ATAC ($\mathcal Y$, orange) state spaces. The corresponding cell populations are represented as densities $\rho_t$ and $\xi_t$ and growth are represented as scalar fields $g_t$ and $h_t$, in $\mathcal{X}$ and $\mathcal{Y}$ respectively.
    \textbf{b,} Left, a trajectory inferred using only RNA data could follow an incorrect path or reach an incorrect terminal state when mapped to ATAC space (red). Right, COATI uses both modalities to jointly constrain trajectory inference.
    \textbf{c,} A neural ODE models RNA trajectories, and a pretrained and fixed FiLM map generates the corresponding ATAC trajectories. Paired trajectories share mass weights. The training objective combines transport and growth costs with reference-matching, manifold, and mass-matching penalties in both spaces.
    \textbf{d,} COATI predicts cell-state distributions at held-out stages and traces developmental lineages across RNA and ATAC. We also use the inferred trajectories to relate population mass changes to developmental programs and examine how chromatin accessibility and gene expression change over time along specific lineages.}
    \label{fig:coati-overview}
\end{figure}

Let \(\rho_t\) and \(\xi_t\) denote population densities in the reduced transcriptomic (RNA) and chromatin-accessibility (ATAC) state spaces \(\mathcal X\) and \(\mathcal Y\), with observations \(\widehat\rho_k\) and \(\widehat\xi_k\) at times \(0=t_0<\cdots<t_K=T\) (Fig.~\ref{fig:coati-overview}a). Given a time-dependent cross-modality map \(\mathbf T_t:\mathcal X\to\mathcal Y\), UnSyncOT jointly minimizes transport and growth costs:

\begin{equation}\label{original_two_unbalanced}
\begin{aligned}
&\inf_{(\rho_t,u_t, g_t)}
\frac{1}{2}\int_0^T 
\left[
c_x\int_{\mathcal X}
\left(\alpha\|u_t\|^2+\beta g_t^2\right)\rho_t\,dx
+
c_y\int_{\mathcal Y}
\left(\alpha\|v_t\|^2+\beta h_t^2\right)\xi_t\,dy
\right]dt
\\[4pt]
&\text{subject to}\quad
\left\{
\begin{array}{@{}l@{}}
\partial_t\rho_t+\alpha\nabla_x\cdot(\rho_tu_t)
=\beta\rho_tg_t, \ \rho_{t_k} = \widehat{\rho}_{k}, \ u_t\cdot n_x = 0 \text{\ on }\partial\mathcal X;\\[2pt]
\partial_t\xi_t+\alpha\nabla_y\cdot(\xi_tv_t)
=\beta\xi_th_t, \ \ \xi_{t_k} = \widehat{\xi}_{k}, \ v_t\cdot n_y = 0 \text{\ on }\partial\mathcal Y;\\[2pt]
\xi_t=(\mathbf{T}_t)_{\#}\rho_t,\quad
h_t(\mathbf T_t(x))=g_t(x);\\[2pt]
v_t(\mathbf T_t(x))=\nabla_x\mathbf T_t(x)u_t(x)+\alpha^{-1}\partial_t\mathbf T_t(x).
\end{array}
\right.
\end{aligned}
\end{equation}
Here, \((u_t,g_t)\) and \((v_t,h_t)\) are the primary and secondary velocity and growth fields, respectively. The modality weights satisfy \(c_x,c_y\geq0\) and \(c_x+c_y=1\), while \(\alpha>0\) and \(\beta\geq0\) control transport and growth. The constraints enforce population evolution, observation matching and synchronization through \(\mathbf T_t\). Thus, both RNA and ATAC observations constrain the inferred dynamics. The detailed formulation and numerical implementation are provided in \hyperref[sec:methods]{Methods}. Trajectories fitted to RNA observations alone may not reproduce the observed ATAC distributions after mapping through $\mathbf T_t$. COATI instead learns the trajectories jointly from RNA and ATAC observations, with both modalities constraining the dynamics during training (Fig.~\ref{fig:coati-overview}b).

By Theorem \ref{theorem:UnSyncOT_reformulation} in \hyperref[sec:methods]{Methods} (also see \cite{cang2026unsyncot}), given the cross-modality map, solving the {UnSyncOT} problem (\ref{original_two_unbalanced}) only requires explicitly estimating the RNA-space velocity and growth fields. In COATI, we parameterize them as neural networks, $u_t\approx\mathrm{NN}_{\theta}(x,t)$ and $g_t\approx\mathrm{NN}_{\phi}(x,t)$. The cross-modality map $\mathbf{T}_t(x)$ is a pretrained and fixed neural network, parameterized as $\mathbf{T}^{\eta}_t(x)$. 
For numerical computation, we represent the evolving cell populations using weighted particles. The learned RNA velocity and growth fields govern particle motion and mass changes, respectively, and the cross-modality map $\mathbf{T}_t$ determines the corresponding ATAC states.
The corresponding particles in RNA space and ATAC space share the same mass weights, and the losses are evaluated in both spaces (Fig.~\ref{fig:coati-overview}c). We use mini-batches of particles to efficiently estimate these costs and replace exact marginal constraints with distribution and mass-matching penalties. 
\hyperref[sec:methods]{Methods} and Supplementary Note 5 derive the joint action of the paired paths for differentiable time-dependent maps. Together, the inferred paired trajectories allow us to investigate developmental transitions, changes in cell-population mass, and the timing of chromatin accessibility changes relative to gene expression (Fig.~\ref{fig:coati-overview}d).

\FloatBarrier

\subsection*{COATI reconstructs coherent population dynamics across RNA and ATAC spaces}
\label{sec:benchmark}

\begin{figure}[!htbp]
    \centering
\includegraphics[width=0.975\textwidth,keepaspectratio]{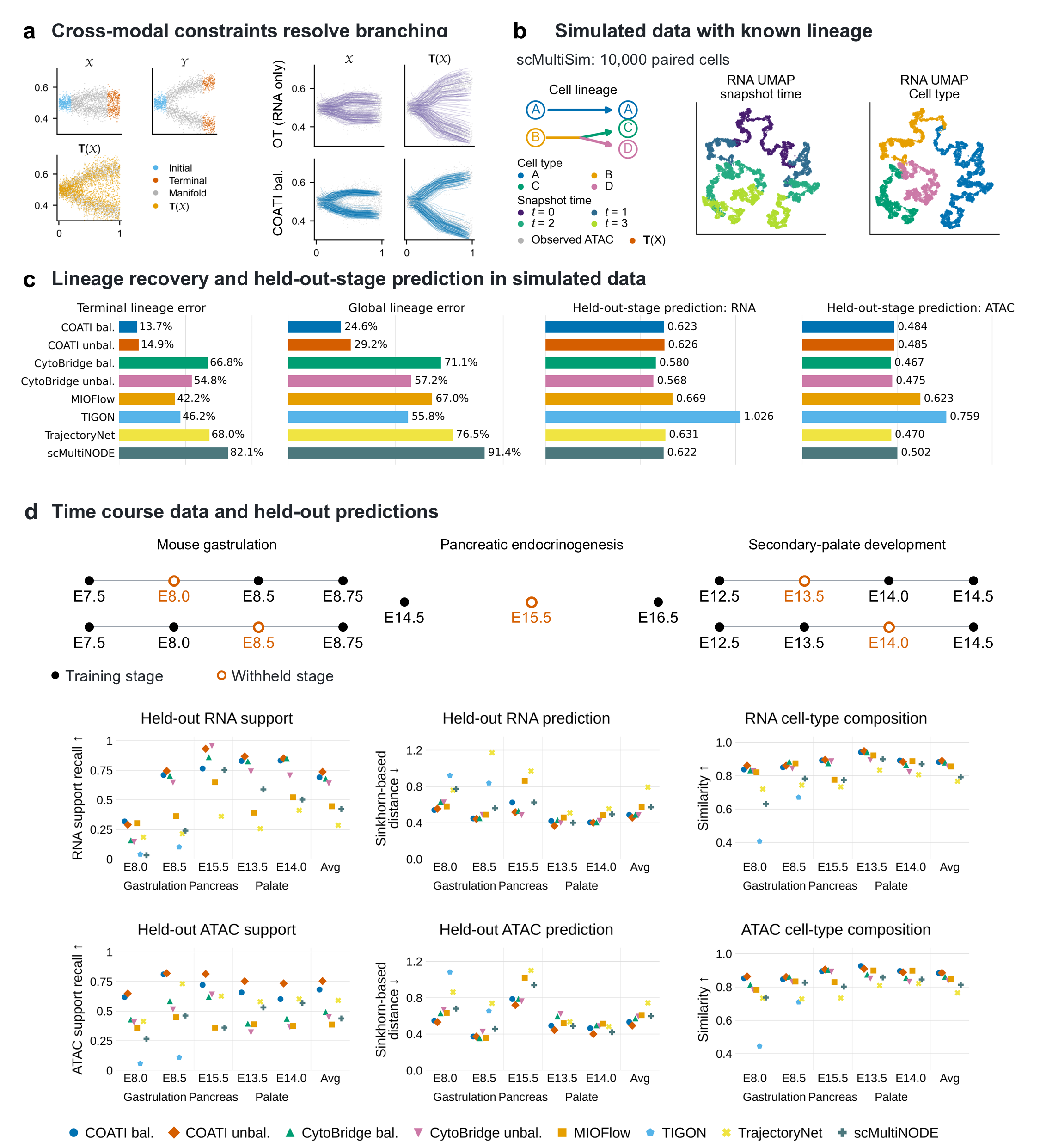}
\caption{\textbf{Trajectory reconstruction and benchmarking of COATI.}
\textbf{a,} Synthetic branching example. Left, simulated populations in the primary space $\mathcal X$ and secondary space $\mathcal Y$, together with the mapped states $\mathbf T(\mathcal X)$. Right, trajectories inferred by RNA-only OT and balanced COATI in $\mathcal X$ and after mapping through $\mathbf T$. Grey points show the reference data manifold.
\textbf{b,} scMultiSim simulated paired RNA-ATAC data from known lineage trees.
\textbf{c,} Terminal and global lineage errors for models trained on all four simulated snapshots (left), and mean RNA and ATAC discrepancies in separate experiments withholding an intermediate snapshot (right). lineage errors use the known ancestry of the simulated cells.
\textbf{d,} Leave-one-stage-out experiments in mouse gastrulation, pancreatic endocrinogenesis, and secondary-palate development. Orange open circles mark withheld stages. The six plots compare local support, Sinkhorn-based distance, and cell-type composition similarity in RNA (top) and ATAC (bottom), across five held-out stages and their averages.
}
\label{fig:benchmark-main}
\end{figure}

We first illustrate how information from a second modality changes the inferred trajectories. We construct a biologically inspired synthetic example in which chromatin accessibility diverges before clear differences emerge in gene expression during progenitor differentiation. RNA-only dynamical OT produces paths that leave the observed ATAC branches after mapping through $\mathbf{T}_t$. COATI instead follows the branching structure in both spaces (Fig.~\ref{fig:benchmark-main}a). We also illustrate COATI in an unbalanced setting with a Gaussian bump example, in which COATI directs trajectories around the bump in the secondary space (Supplementary Fig. 1). We set the terminal mass to twice the initial mass in both examples and vary $\beta$, which controls the balance between transport and local mass changes. Larger $\beta$ favors mass creation and removal over transport. The branching structure remains consistent across the tested $\beta$ values, despite changes in the inferred growth rates (Supplementary Fig. 1).

We next evaluate COATI on a paired RNA-ATAC dataset of 10{,}000 cells simulated using scMultiSim with a predefined lineage tree~\cite{scmultisim}. We divide the cells into four pseudotime-ordered snapshots, each normalized to unit mass. The known lineage tree contains two starting populations where $\mathcal A$ remains $\mathcal A$, and $\mathcal B$ gives rise to $\mathcal C$ and $\mathcal D$ (Fig. \ref{fig:benchmark-main}b). We assess lineage recovery using models trained on all four snapshots, and population reconstruction in separate experiments withholding $t=1$ or $t=2$. To assess lineage recovery, we use a terminal lineage error which measures whether trajectories end in cell types inconsistent with their ancestral cell type, and global lineage error further accounts for trajectories that enter incorrect intermediate lineages (\hyperref[sec:methods]{Methods}). We compare COATI with TrajectoryNet, MIOFlow, TIGON, scMultiNODE, and CytoBridge. Both COATI variants achieve the lowest terminal and global lineage errors, while their mean held-out RNA and ATAC reconstruction errors are comparable to those of several other methods (Fig.~\ref{fig:benchmark-main}c). Terminal distributional fits are also comparable among several methods (Supplementary Fig. 2). To assess the contribution of synchronization, we compare COATI with matched RNA-only OT baselines. Both COATI variants reduce global lineage error across all tested synchronization weights (Supplementary Fig. 3a) and terminal lineage error with each of the five tested map architectures (Supplementary Fig. 3b). The time-conditioned FiLM map gives the lowest terminal error and is used in subsequent analyses. This reduction in error relative to the RNA-only baselines is also observed in RNA and ATAC reconstruction at both withheld times (Supplementary Fig. 3c).

We then evaluate population reconstruction using real time course data of paired RNA-ATAC of mouse gastrulation~\cite{argelaguet2022decoding}, pancreatic endocrinogenesis~\cite{klein2025mapping} and secondary palate development~\cite{yan2024single}. We withhold E8.0 or E8.5 in gastrulation, E15.5 in pancreas, and E13.5 or E14.0 in palate, giving five separate prediction tasks (Fig.~\ref{fig:benchmark-main}d). For COATI, the cross-modality map is also trained on data excluding the held-out stage. All methods are evaluated in common RNA and ATAC embedding spaces. We assess three aspects of reconstruction including local support that measures how well predictions cover neighborhoods of observed cells, distributional discrepancy that measures agreement between the predicted and observed cells, and cell type composition similarity that measures agreement in cell type proportions (\hyperref[sec:methods]{Methods}). Across the five held-out stages, unbalanced COATI achieves the best average scores for all three metrics in both modalities (Fig.~\ref{fig:benchmark-main}d and Supplementary Figs. 4–6). Compared with matched RNA-only OT baselines, both COATI variants substantially increase ATAC support, with smaller reductions in RNA support. Cell type composition similarity improves in RNA and generally improves in ATAC, while distributional distances remain similar in both modalities (Supplementary Fig. 7a). To assess sensitivity to training randomness, we repeat the E13.5 palate experiment with different random seeds. Both variants yield consistent support, distributional distance and composition scores across these runs in both modalities (Supplementary Fig. 7b).
Further numerical experiments demonstrate the expected trade-off between RNA and ATAC kinetic energies controlled by the synchronization weight $c_y$, and support the quality of the cross-modality map $\mathbf{T}_t$ in preserving ATAC structure and cell-type correspondence (Supplementary Figs. 8--11).

\FloatBarrier

\subsection*{Reconstructing gastrulation states and posterior axial differentiation}
\label{sec:gastrulation-results}

\begin{figure}[!htbp]
\centering
\includegraphics[width=0.67\textwidth,keepaspectratio]{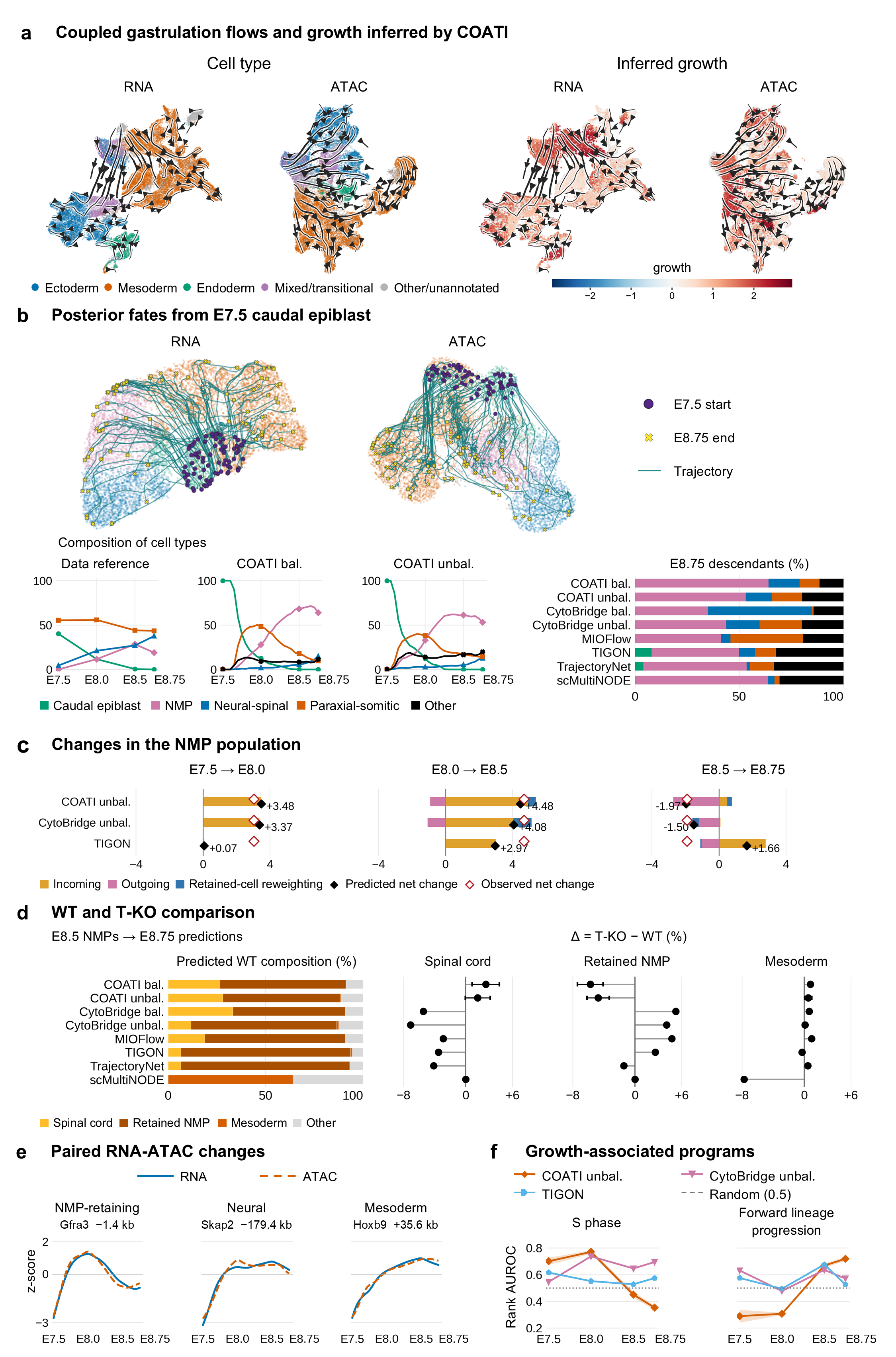}
\caption{\textbf{COATI reconstructs posterior cell fates and molecular changes during gastrulation.}
\textbf{a,} RNA and ATAC flows inferred by COATI using all four stages from E7.5 to E8.75. Cells are colored by germ-layer identity (left) or inferred growth (right).
\textbf{b,} RNA and ATAC trajectories starting from NMP-related subgroups at E7.5. Below are the observed composition of NMP-associated cell types, predicted composition along COATI trajectories starting from caudal epiblast at E7.5, and E8.75 fate compositions across methods.
\textbf{c,} Changes in the NMP proportion between successive stages, separated into incoming mass, outgoing mass, and changes in the relative weights of cells remaining NMPs.  Black diamonds show predicted net changes and open red diamonds show observed changes, in percentage points.
\textbf{d,} Predicted E8.75 fates of measured E8.5 wild-type (WT) and Brachyury (\textit{T})-knockout (\textit{T}-KO) NMPs. Left, predicted WT composition. Right, changes in spinal-cord, retained-NMP, and mesoderm proportions (\textit{T}-KO minus WT, in percentage points). scMultiNODE uses its decoded ATAC states. For COATI, whiskers show standard deviations across nine synchronization weights.
\textbf{e,} Standardized RNA expression (solid blue) and ATAC peak accessibility (dashed orange) along NMP-retaining, neural, and mesoderm trajectories for peak-gene pairs involving \textit{Gfra3}, \textit{Skap2}, and \textit{Hoxb9}. Peak positions relative to each gene's transcription start site are shown.
\textbf{f,} AUROC for S-phase and genes associated with lineage progression ranked by their association with inferred growth across stages. Values above 0.5 indicate that genes in the program tend to rank above other genes.
}
\label{fig:gastrulation}
\end{figure}

We return to the mouse gastrulation atlas to examine developmental fates and molecular changes along the inferred trajectories~\cite{argelaguet2022decoding}. Jointly modeling all four developmental stages, from E7.5 to E8.75, yields corresponding RNA and ATAC flows and inferred growth fields (Fig.~\ref{fig:gastrulation}a). We first assess whether the inferred trajectories preserve developmental lineage relationships by measuring the fraction of cells assigned to a different germ layer from their starting population. Both balanced and unbalanced COATI reduce these off-lineage assignments compared to RNA-only OT baselines (Supplementary Fig. 12a,b). We further examined rostral-neuroectoderm cells with chromatin accessibility characteristic of neural identity, which are expected to follow neural rather than blood/endothelial fates. At held-out E8.5, balanced COATI assigns only 1.1\% of their predicted descendants to blood/endothelial states, compared with 17.1\% for RNA-only OT results (Supplementary Fig. 12c,d).

We next follow trajectories from E7.5 caudal-epiblast cells through neuromesodermal progenitor (NMP) states toward neural–spinal and paraxial–somitic fates. Both COATI variants retain these two branches, whereas several other methods assign few descendants to one of them (Fig.~\ref{fig:gastrulation}b and Supplementary Fig. 13). The predicted NMP fraction first increases and then declines, resembling the trend in the observed atlas. We also test whether the predicted descendants reproduce the molecular changes associated with NMP differentiation by comparing their RNA and ATAC differences from NMPs with the corresponding observed differences. COATI gives the closest agreement at held-out E8.0 and E8.5 across both branches and modalities among methods with sufficient predicted descendants for this comparison (Supplementary Fig. 14). To understand what drives the changes in the NMP proportion, we track the full E7.5 population and separate contributions into three types including mass entering the NMP state, mass leaving it, and changes in the relative weights of cells remaining NMPs (Methods). Unbalanced COATI attributes the early increase mainly to incoming transitions and the later decline mainly to outgoing differentiation (Fig.~\ref{fig:gastrulation}c). From E8.5 to E8.75, it predicts a decline of 1.97\%, close to the observed decline of 1.92\%.

We then test the learned dynamics on experimentally perturbed starting states. In the observed E8.5 samples, Brachyury (\textit{T}) knockout (\textit{T}-KO) strongly reduces the somitic-mesoderm fraction while preserving the spinal-cord fraction. The RNA-to-ATAC map trained without perturbation data produced cell-type assignments that agreed with those from the paired observed ATAC profiles for 83.9\% of wild-type (WT) and 78.7\% of \textit{T}-KO NMP cells (Supplementary Fig. 15a,b).
We evolve the measured WT and \textit{T}-KO NMPs from E8.5 to E8.75 using models trained without perturbation samples. Both COATI variants predict a small increase in the spinal-cord fraction in \textit{T}-KO relative to WT, whereas CytoBridge, MIOFlow, TIGON and TrajectoryNet predict decreases and scMultiNODE predicts no spinal-cord descendants in either condition (Fig.~\ref{fig:gastrulation}d). Across all models, only COATI's RNA and ATAC readouts agree on the mean directions of change in both the spinal-cord and retained-NMP proportions (Supplementary Fig. 15c,d).

The paired trajectories also allow us to examine how gene expression and chromatin accessibility change together within each lineage. We first screened the observed paired RNA–ATAC data for peak–gene associations, requiring significant positive correlations after adjusting for stage and cell type in both replicate series (\hyperref[sec:methods]{Methods}). This yielded 315 candidate pairs, which we ranked by the similarity of their temporal changes along COATI trajectories.
The highest-ranked pairs involve \textit{Gfra3}, \textit{Skap2} and \textit{Hoxb9} in NMP-retaining, neural and mesoderm trajectories, respectively (Fig.~\ref{fig:gastrulation}e). \textit{Gfra3} expression and accessibility increase and decrease together, and the \textit{Skap2} and \textit{Hoxb9} profiles increase along their respective lineages. Candidate pairs showed higher temporal similarity than alternative pairings of the same genes and peaks (Supplementary Fig. 16a). All three displayed peaks overlapped ENCODE candidate cis-regulatory elements (Supplementary Fig. 16b).

Finally, we examine whether inferred growth is associated with cell-cycle activity or differentiation. We rank genes by their partial Spearman correlation with growth and assess whether genes in each program rank above other genes using AUROC (\hyperref[sec:methods]{Methods}). Higher AUROCs indicate that genes in the program tend to have higher signed growth correlations than background genes. In unbalanced COATI, S-phase genes rank highly at E7.5–E8.0, and forward-lineage-progression genes rank higher at E8.5–E8.75 (Fig.~\ref{fig:gastrulation}f). At E8.5, growth is positively associated with maturation programs in most of the examined cell types, while somitic mesoderm shows a stronger association with S-phase activity (Supplementary Fig. 16c,d). These results link inferred growth to different biological programs across stages and cell types.

\FloatBarrier

\subsection*{Resolving endocrine lineage transitions with chromatin information}
\label{sec:pancreas}
\begin{figure}[!htbp]
\centering
\includegraphics[width=0.825\textwidth,keepaspectratio]{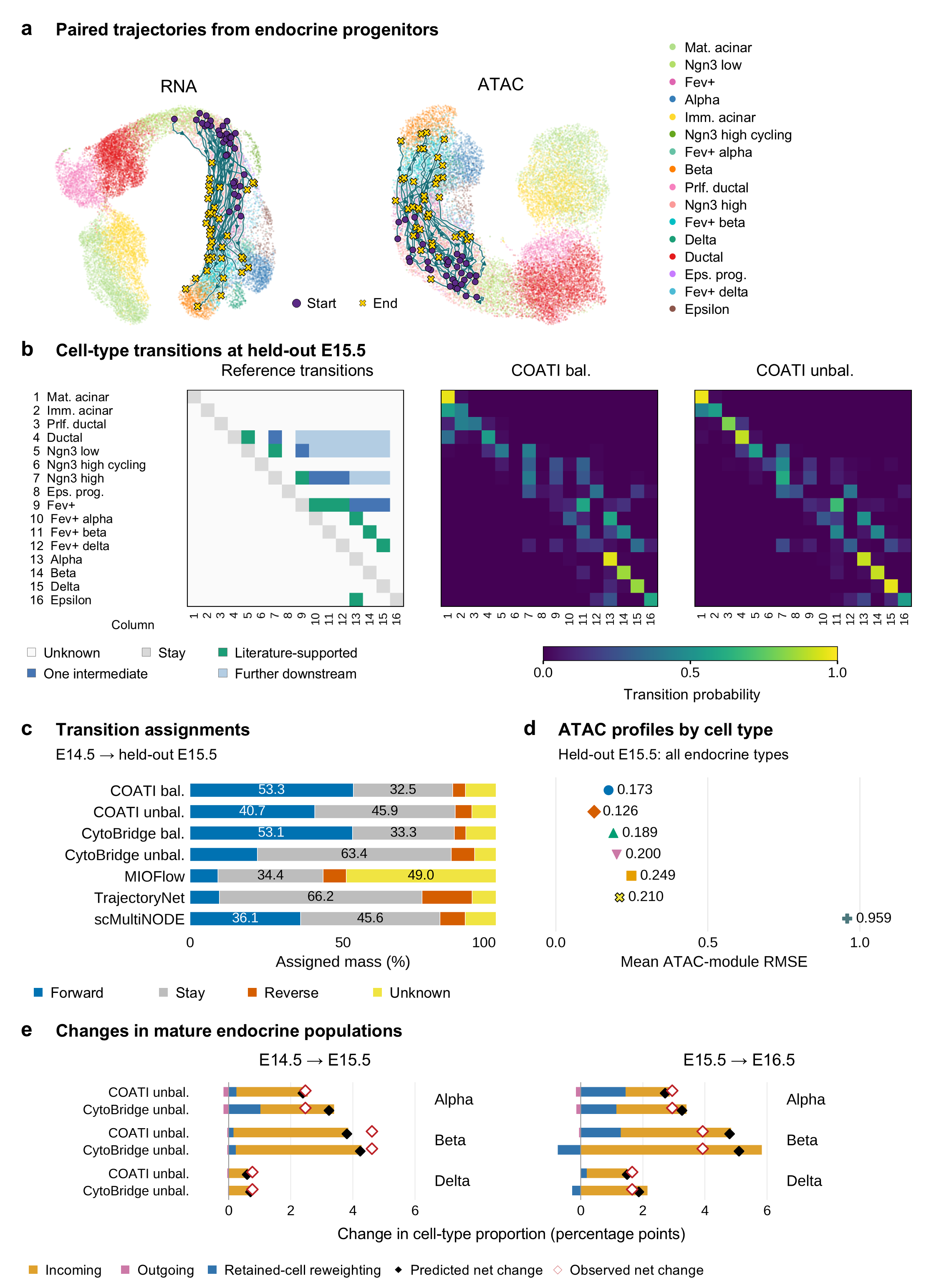}
\caption{\textbf{Endocrine lineage transitions and chromatin states during pancreas development.}
\textbf{a,} Corresponding RNA and ATAC trajectories from E14.5 Ngn3-high progenitors to E16.5, inferred by COATI using all three stages. Background points show observed cell types.
\textbf{b,} Literature-supported reference transitions and predicted E14.5-to-E15.5 transition matrices from COATI, with E15.5 withheld from training. Rows represent starting cell types and columns represent predicted cell types.
\textbf{c,} Fractions of predicted mass assigned to forward, unchanged, reverse, or unclassified transitions using the reference in \textbf{b}. Unclassified transitions have no assigned direction in that reference.
\textbf{d,} Error in ATAC profiles for predicted endocrine cell types at held-out E15.5. Root-mean-square error (RMSE) is calculated using four accessibility gene programs associated with alpha, beta, delta, and epsilon cells and averaged across cell types weighted by observed cell counts.
\textbf{e,} Changes in alpha, beta, and delta cell proportions in models trained on all three stages, separated into incoming mass, outgoing mass, and reweighting of cells retaining the same identity.
}
\label{fig:pancreas}
\end{figure}

We next examine pancreatic endocrine differentiation using paired RNA and ATAC profiles sampled at E14.5, E15.5 and E16.5~\cite{klein2025mapping}. COATI reconstructed trajectories from E14.5 Ngn3-high progenitors to E16.5. The inferred trajectories move from the progenitor region toward downstream endocrine states in both RNA and ATAC spaces (Fig. \ref{fig:pancreas}a). 
We then evaluate whether the inferred trajectories recover known developmental relationships when E15.5 is withheld from training. We compare the predicted E14.5-to-E15.5 transitions with a reference constructed from published developmental and lineage-tracing studies~\cite{shih2012notch,bastidasPonce2019comprehensive,byrnes2018lineage,arnes2012ghrelin,yu2021sequential} (Fig.~\ref{fig:pancreas}b and Supplementary Fig. 17). We classify transitions as forward, unchanged, reverse or unclassified according to this reference (\hyperref[sec:methods]{Methods}). COATI assigns most predicted mass to forward developmental transitions or unchanged cell states, with few reverse transitions (Fig.~\ref{fig:pancreas}c).
Next, we examine whether the predicted endocrine cell types have the expected chromatin-accessibility profiles. For each cell type, we compare four ATAC modules associated with alpha, beta, delta and epsilon cells against the corresponding observed E15.5 profiles. Unbalanced COATI achieves the lowest overall error, with a mean RMSE weighted by observed cell counts of 0.126, followed by balanced COATI at 0.173 (Fig.~\ref{fig:pancreas}d). COATI also gives the lowest errors for progenitor and mature endocrine populations, indicating that the reconstructed chromatin profiles preserve features of both early and differentiated cell states (Supplementary Fig. 18).

We further examine whether the models reproduce changes in endocrine cell-type proportions. In the held-out E15.5 experiment, COATI captures the observed decrease in Ngn3-high progenitors and increases in alpha, beta and delta cells (Supplementary Fig. 19a). To understand the changes in mature endocrine populations, we use the models trained on all three stages and separate the contributions from incoming mass, outgoing mass and reweighting of cells retaining their identities. From E15.5 to E16.5, COATI closely reproduces the observed increases in alpha, beta and delta proportions (Fig.  \ref{fig:pancreas}e). For beta and delta cells, both incoming transitions and increased relative weights of cells retaining these identities contribute to the predicted increase in their sampled proportions. COATI also infers positive mean growth rates for mature endocrine cells and negative rates for Ngn3 progenitors, consistent with the increasing representation of differentiated endocrine populations (Supplementary Fig. 19b).

\FloatBarrier

\subsection*{Preserving anterior--posterior regulatory structure during palate development}
\label{sec:palate}
\begin{figure}[!htbp]
\centering
\includegraphics[width=0.95\textwidth,keepaspectratio]{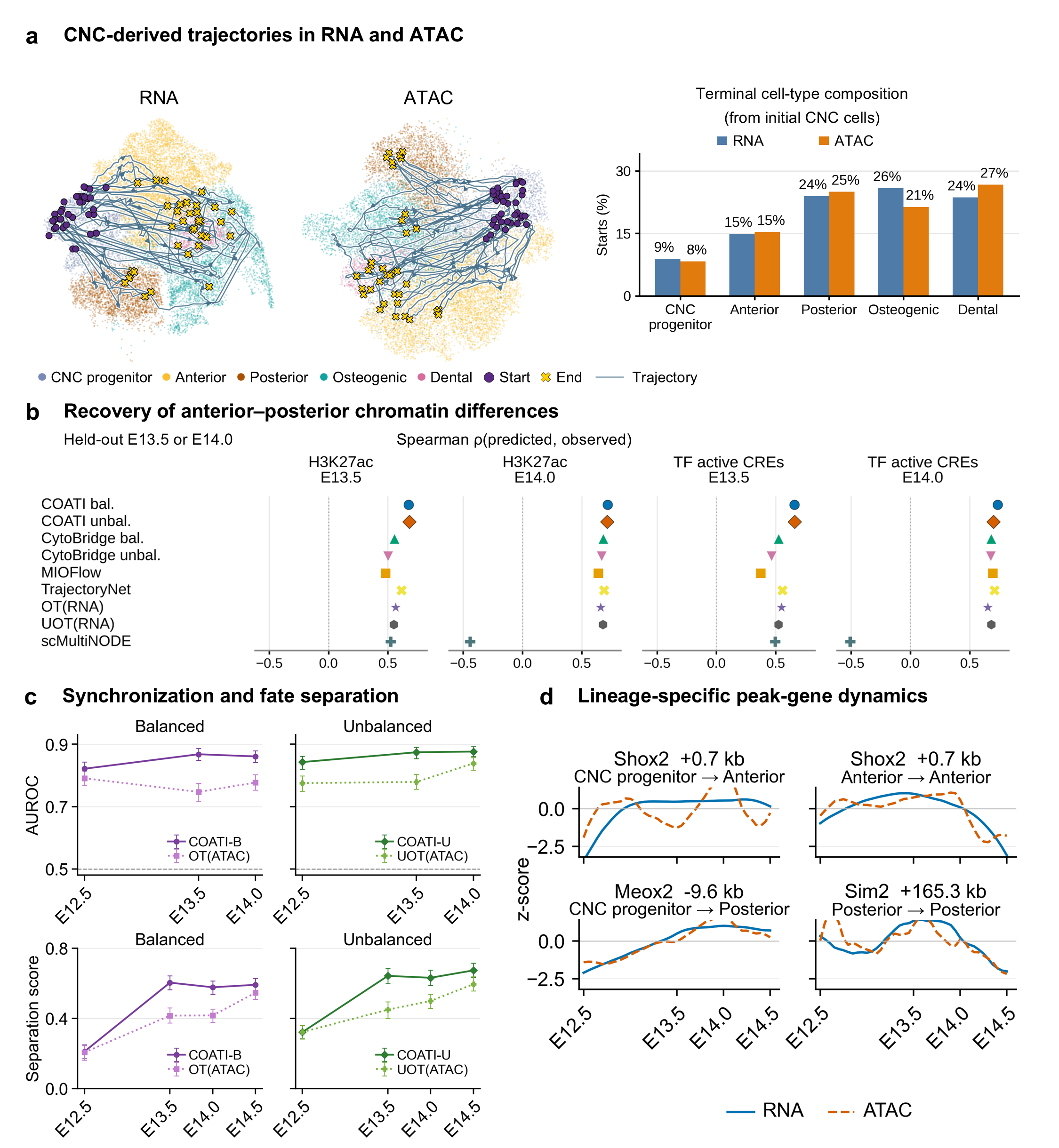}
\caption{\textbf{Palate trajectories and anterior--posterior chromatin patterns.}
\textbf{a,} Corresponding RNA and ATAC trajectories from E12.5 cranial neural crest (CNC) progenitors to E14.5, inferred by COATI using all four stages. Bars show RNA- and ATAC-based terminal cell-type compositions for all 1,048 CNC starting cells.
\textbf{b,} Spearman correlations between predicted and observed anterior-minus-posterior accessibility differences at held-out E13.5 and E14.0, measured across H3K27ac regions and transcription-factor-bound active regulatory elements defined from published data. Higher correlations indicate better recovery of the observed regional differences.
\textbf{c,} H3K27ac-based discrimination of predicted anterior and posterior fates along trajectories. Top, AUROC through E14.0; bottom, separation of the regional accessibility scores through E14.5. Error bars show 95\% particle-bootstrap intervals from 500 resamples.
\textbf{d,} Standardized RNA expression (solid blue) and ATAC peak accessibility (dashed orange) along CNC-to-anterior, anterior-retaining, CNC-to-posterior, and posterior-retaining trajectories. The displayed peak--gene pairs involve \textit{Shox2}, \textit{Meox2}, and \textit{Sim2}. Labels describe peak positions relative to the gene transcription start sites.
}
\label{fig:palate}
\end{figure}
Here, we study anterior–posterior regulatory programs during mouse secondary-palate development using paired RNA and ATAC profiles from E12.5 to E14.5~\cite{yan2024single}. During this period, cranial neural crest (CNC)-derived cells differentiate into anterior and posterior palatal mesenchyme, as well as osteogenic and dental populations. Using all sampled stages, COATI infers corresponding RNA and ATAC trajectories from E12.5 CNC progenitors to these downstream states (Fig.~\ref{fig:palate}a and Supplementary Fig. 20). 

Anterior and posterior palate development involves distinct regulatory programs where \textit{Shox2} is associated with the anterior palate, and \textit{Meox2} is associated with the posterior palate~\cite{yan2024single,xu2019shox2}. We ask whether the inferred chromatin states preserve these regional differences when data from an intermediate stage are withheld from training. Using H3K27ac regions and active regulatory elements bound by the SHOX2 or MEOX2 proteins, as identified in published experiments, we compared predicted and observed anterior-minus-posterior accessibility contrasts. COATI recovers these differences at held-out E13.5 and E14.0 across both sets of regulatory regions (Fig.~\ref{fig:palate}b). The regional patterns are also recovered when all stages are used for training (Supplementary Fig. 21a), consistent with the distinct chromatin programs associated with anterior and posterior palate identity. To investigate the negative  correlations for scMultiNODE (Supplementary Fig. 21a), we examined its individual decoded states on RNA and ATAC. Although the decoded distributions resembled the observations for both modalities, cell-type preservation was poor, particularly in ATAC (Supplementary Fig. 21b,c).
COATI also recovers chromatin variation within individual cell types, including less abundant ATAC substates, at both held-out stages (Supplementary Fig. 22).

We also ask whether early chromatin accessibility is associated with the anterior or posterior fate of the inferred trajectories. For trajectories starting from E12.5 CNC progenitors, H3K27ac-based regional scores distinguish cells assigned to anterior and posterior states at E14.5, with greater separation at later stages (Fig.~\ref{fig:palate}c). Synchronization with RNA increases this separation relative to ATAC-only transport. A similar improvement is seen with TF-bound regulatory regions (Supplementary Fig. 23). These results are consistent with regional chromatin programs becoming established as palatal progenitors differentiate.

Lastly, we follow expression and accessibility of candidate peak–gene pairs along the inferred lineages. Candidate pairs associated with anterior and posterior genes were screened from the observed paired data and ranked by the similarity of their temporal changes along COATI trajectories (Methods). The highest-ranked examples include the same \textit{Shox2}-associated peak along both CNC-to-anterior and trajectories retaining an anterior identity, a \textit{Meox2} pair along CNC-to-posterior trajectories, and a \textit{Sim2} pair along posterior-retaining trajectories (Fig.~\ref{fig:palate}d). These genes have established anterior or posterior expression patterns during palate development~\cite{yan2024single}. The \textit{Shox2}- and \textit{Sim2}-associated peaks overlap region-matched H3K27ac signals. Additional examples further show corresponding RNA and ATAC changes along these developmental paths (Supplementary Fig. 24).

\FloatBarrier

\subsection*{Reconstructing transcriptional and chromatin dynamics in cerebral organoids}
\label{sec:human-cerebral}

\begin{figure}[!htbp]
\centering
\includegraphics[width=0.72\textwidth,keepaspectratio]{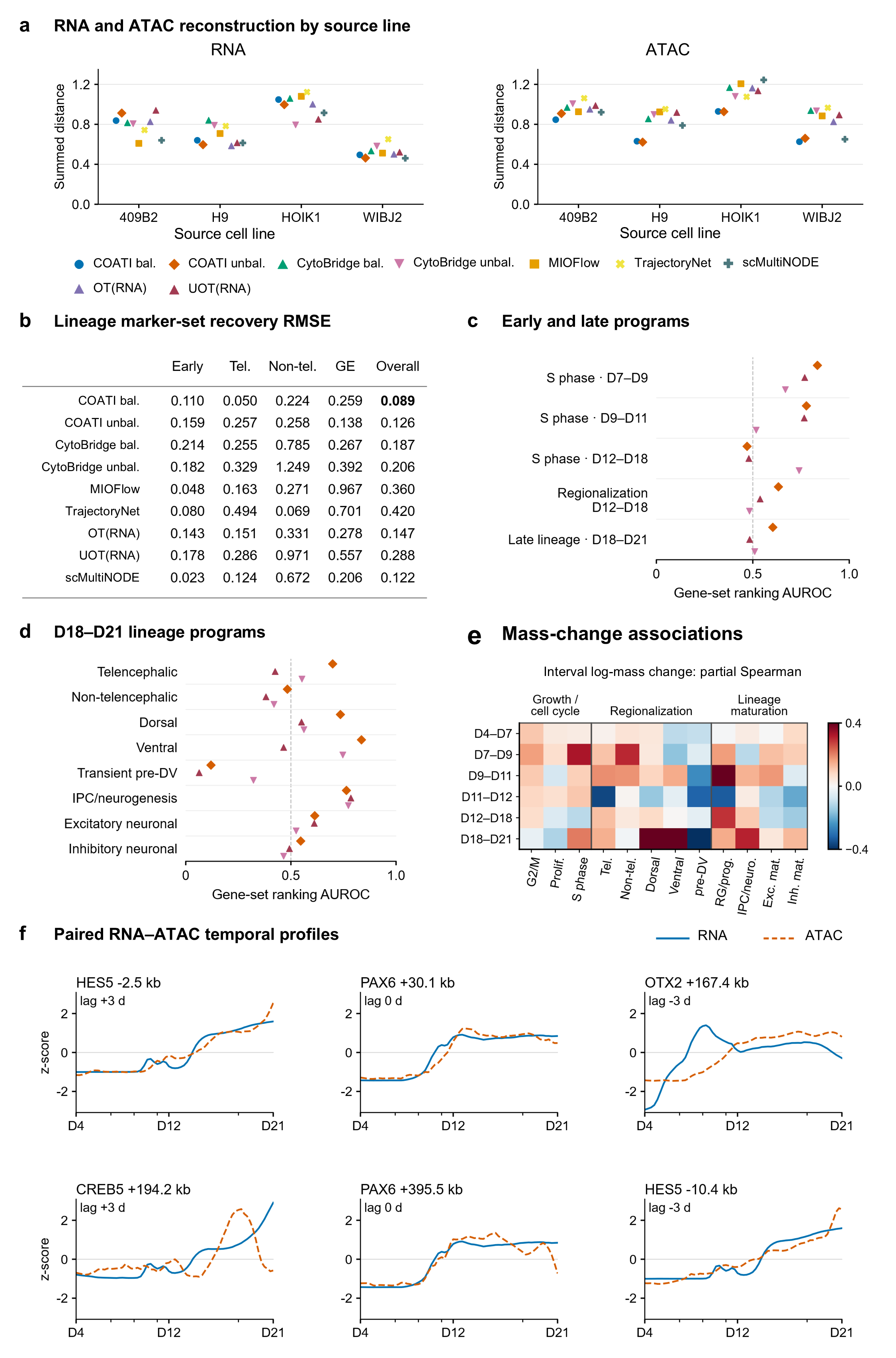}
\caption{\textbf{Neural development and paired RNA--ATAC dynamics in human cerebral organoids.}
All panels use models trained on all seven stages of computationally paired RNA--ATAC metacells.
\textbf{a,} RNA and ATAC distribution discrepancies for the four D4 source stem-cell lines, summed over the six later sampled stages. Lower values indicate closer agreement with observed distributions.
\textbf{b,} Root-mean-square error (RMSE) in lineage marker-gene expression across early, telencephalic, non-telencephalic, and ganglionic-eminence populations. The overall score summarizes groups defined by stage and lineage evaluated for all methods.
\textbf{c,} Gene-set AUROC for S-phase, regionalization, and late-lineage genes, ranked by the association of starting expression with subsequent accumulated log-mass change over the indicated intervals.
\textbf{d,} Gene-set AUROC for eight neural-development programs over D18--D21, using the same mass-change ranking.
\textbf{e,} Partial Spearman correlations between gene-program scores and interval log-mass changes for unbalanced COATI.
\textbf{f,} Standardized RNA expression (solid blue) and ATAC peak accessibility (dashed orange) along COATI trajectories for six candidate peak-gene pairs involving \textit{HES5}, \textit{PAX6}, \textit{OTX2}, and \textit{CREB5}. Labels show distances from peaks to the corresponding transcription start sites and the reported temporal shifts.
}
\label{fig:cerebral}
\end{figure}

In the last section, we study neural induction and regionalization in human cerebral organoids using a time-course dataset in which RNA and ATAC were measured in separate cells~\cite{fleck2023inferring}. We use computationally paired RNA-ATAC metacells for COATI analysis. The data span seven developmental stages and four stem-cell lines. We train COATI using all stages, starting from D4, without using labels indicating the stem cell line of origin to constrain the trajectories. COATI reconstructed the RNA and ATAC distributions of the individual source lines, with particularly good agreement in ATAC space (Fig.~\ref{fig:cerebral}a). The RNA-to-ATAC map also preserved information about cell type and source line (Supplementary Fig. 25a).

To examine whether the inferred populations preserve neural identities, we compared their lineage-associated gene expression scores with observations from the same stage and lineage, using marker sets from the original study. COATI recovered these expression patterns across early, telencephalic, non-telencephalic and ganglionic-eminence groups, with the lowest overall error among the methods evaluated (Fig.~\ref{fig:cerebral}b). 
Unbalanced COATI also infers changes in population mass along trajectories. To understand their biological associations, we ranked genes by their signed correlations with subsequent accumulated log-mass change and assessed cell-cycle and developmental gene sets using AUROC (Methods). S-phase genes rank highly during D7–D9 and D9–D11, which is consistent with early progenitor expansion. During D12–D18, this association weakens with regionalization genes ranking higher (Fig.~\ref{fig:cerebral}c). From D18 to D21, telencephalic, dorsal and ventral gene sets show strong associations with inferred mass changes, consistent with the emergence of regional neural identities (Fig.~\ref{fig:cerebral}d). Correlations with gene-set scores show a similar shift, with weaker G2/M associations and stronger positive associations with dorsal and ventral programs at later stages (Fig.~\ref{fig:cerebral}e).

Finally, we use the paired trajectories to examine how gene expression and chromatin accessibility change together during neural development. We first evaluate eight previously identified peak–gene pairs involving \textit{HES4}, \textit{HES5} and \textit{CREB5}, whose expression and accessibility decreased following \textit{GLI3} knockout in the original study~\cite{fleck2023inferring}. COATI shows greater temporal RNA–ATAC agreement than RNA-only transport on these pairs, although scMultiNODE achieved the highest mean score (Supplementary Fig. 25b). We also follow six additional candidate peak–gene profiles involving \textit{HES5}, \textit{PAX6}, \textit{OTX2} and \textit{CREB5} along COATI trajectories (Fig.~\ref{fig:cerebral}f). Five of the six peaks overlap GLI3 CUT\&Tag regions, and both \textit{HES5}-associated peaks overlap regions with reduced accessibility following GLI3 knockout~\cite{fleck2023inferring}. These findings connect the inferred RNA and ATAC dynamics with regulatory regions implicated in neural development.

\FloatBarrier

\section*{Discussion}
COATI extends synchronized optimal transport to time-dependent cross-modality maps and provides a neural network framework for jointly reconstructing corresponding RNA and ATAC trajectories in separate state spaces. Across synthetic benchmarks and developmental datasets, joint trajectory learning improved the recovery of chromatin-state heterogeneity and developmental lineage relationships, as well as reconstruction of held-out cell populations. Comparisons with RNA-only models showed that ATAC observations can help determine the inferred trajectories.

COATI and scMultiNODE take different approaches to modeling RNA and ATAC dynamics. scMultiNODE learns trajectories in a shared latent space, whereas COATI reconstructs corresponding trajectories in separate RNA and ATAC spaces. By evaluating transport costs and agreement with observations in both spaces, COATI allows each modality to contribute its own information about developmental transitions. This is particularly useful when chromatin accessibility distinguishes cell states that are similar in gene expression. The strong temporal agreement between RNA and ATAC changes achieved by scMultiNODE in the cerebral organoid analysis highlights the value of shared representations. An interesting future direction would be to combine shared representation learning with synchronized transport in separate spaces, allowing cross-modality relationships and developmental trajectories to be refined jointly.

Leave-one-stage-out evaluation in the three mouse datasets tested the reconstruction of populations at unobserved developmental stages. These predictions can be more challenging when observations are widely spaced or when early lineage differences are subtle. Moreover, agreement with a held-out cell distribution does not establish whether the inferred trajectories follow the correct developmental lineages. We therefore complemented distributional metrics with evaluations of cell-type transitions, developmental constraints and regulatory features. These analyses provided additional evidence that COATI recovered biologically meaningful developmental trajectories.

The corresponding RNA and ATAC trajectories also provide a way to examine how gene expression and chromatin accessibility change along specific developmental lineages. Correlations across observed cells can reflect differences between cell types and changes occurring during differentiation. Following defined source-to-target trajectories allows these patterns to be examined separately and can help identify candidate peak–gene relationships with coordinated temporal changes. Several of these peaks also overlapped regulatory regions identified in previous studies.

The unbalanced formulation further allows changes in population mass along the inferred trajectories. In the gastrulation and cerebral organoid analyses, inferred mass changes were more strongly associated with cell-cycle programs at early stages and with lineage differentiation or regionalization programs at later stages. These associations provide a biological interpretation of the inferred growth fields. When snapshots are normalized to unit mass, the model describes relative changes in population mass rather than direct estimates of cell proliferation or death. When the total population size of the biological system is known and used to set the total mass at each stage, the inferred mass changes can be interpreted more directly, though still approximately, in terms of cell proliferation and death.

Lastly, COATI currently relies on a pretrained RNA-to-ATAC map, and the accuracy of this map affects the inferred trajectories in both modalities. Improving cross-modality mapping and incorporating additional information about population changes could further strengthen the reconstruction. Overall, our results demonstrate the value of jointly modeling transcriptional and chromatin dynamics while preserving the distinct information provided by each modality.

\section*{Methods}
\label{sec:methods}

\subsection*{Notations and Formulations of UnSyncOT}
Let time points $0 = t_0 < \cdots < t_k < \cdots t_{K}=T$ denote the observation times. Let $\mathcal X$ be the RNA state space (respectively $\mathcal Y$ the ATAC state space) with marginal densities $\{\widehat\rho_k\}_{k=0}^K$ (respectively $\{\widehat\xi_k\}_{k=0}^K$). Denote  $\mathbf T_t = \mathbf{T}(x,t):\mathcal X\times[0,T]\to\mathcal Y$ the prescribed continuously differentiable time-dependent cross-modality map. The unbalanced synchronized dynamical optimal transport (UnSyncOT) is formulated as
\begin{equation}\label{eqn:UnSyncOT01}
\begin{aligned}
&\inf_{(\rho_t,u_t, g_t)}
\frac{1}{2}\int_0^T 
\left[
c_x\int_{\mathcal X}
\left(\alpha\|u_t\|^2+\beta g_t^2\right)\rho_t\,dx
+
c_y\int_{\mathcal Y}
\left(\alpha\|v_t\|^2+\beta h_t^2\right)\xi_t\,dy
\right]dt
\\[4pt]
&\text{subject to}\quad
\left\{
\begin{array}{@{}l@{}}
\partial_t\rho_t+\alpha\nabla_x\cdot(\rho_tu_t)
=\beta\rho_tg_t, \ \rho_{t_k} = \widehat{\rho}_{k}, \ k=0,\cdots,K,  \ u_t\cdot n_x = 0 \text{\ on }\partial\mathcal X;\\[2pt]
\partial_t\xi_t+\alpha\nabla_y\cdot(\xi_tv_t)
=\beta\xi_th_t, \ \ \xi_{t_k} = \widehat{\xi}_{k}, \ k=0,\cdots,K, \ v_t\cdot n_y = 0 \text{\ on }\partial\mathcal Y;\\[2pt]
\xi_t=(\mathbf{T}_t)_{\#}\rho_t,\
h_t(\mathbf T_t(x))=g_t(x), \ 
v_t(\mathbf T_t(x))=\nabla_x\mathbf T_t(x)u_t(x)+\alpha^{-1}\partial_t\mathbf T_t(x).
\end{array}
\right.
\end{aligned}
\end{equation}
Here, $\rho_t$ and
$\xi_t$ are time-dependent density trajectories in the two feature spaces, respectively. The velocity and growth rate in $\mathcal X$ (respectively in $\mathcal Y$) are $u_t$ and $g_t$ (respectively $v_t$ and $h_t$). The triplet $(\rho_t, u_t, g_t)$ satisfies the continuity equation with growth in $\mathcal X$, with marginal conditions and a no-flux boundary condition. Similarly, $(\xi_t, v_t, h_t)$ satisfies the corresponding conditions in $\mathcal Y$. The total action is the weighted sum of the actions in $\mathcal X$ and $\mathcal Y$, with nonnegative modality weights $c_x, c_y$ satisfying $c_x+c_y=1$. Parameters $\alpha>0$ and $\beta\geq0$ control the trade-off between mass transport and mass teleportation (namely, creation or destruction). Throughout the paper, we refer to \(\mathcal{X}\) and \(\mathcal{Y}\) as the primary and secondary spaces, respectively. Accordingly, we use the modifiers “primary” and “secondary” to distinguish quantities defined in these spaces. For example, \(u_t\) is the primary velocity, and \(h_t\) is the secondary growth term.

In the feasibility conditions, the pushforward condition $\xi_t = (\mathbf{T}_t)_{\sharp}\rho_t$ maps the RNA population to the ATAC population. The growth relation $h_t(\mathbf T_t(x))=g_t(x)$ ensures that an RNA trajectory $X_t$ and the induced ATAC trajectory $Y_t = \mathbf{T}_t(X_t)$ carry the same mass.  The velocity relation $v_t(\mathbf T_t(x))=\nabla_x\mathbf T_t(x)u_t(x)+\alpha^{-1}\partial_t\mathbf T_t(x)$ follows by applying the chain rule on $Y_t = \mathbf{T}_t(X_t)$, given an RNA trajectory $X_t$. We also assume compatible marginal densities, $\widehat\xi_k=(\mathbf T_{t_{k}})_{\#}\widehat\rho_k$ for $k=0,\cdots,K$. Under these conditions, the joint minimization (\ref{eqn:UnSyncOT01}) is determined by the primary triplet $(\rho_t,u_t,g_t)$, as shown in Theorem \ref{theorem:UnSyncOT_reformulation} below.

Since the RNA and ATAC datasets usually have different scales, the trade-off between transportation and teleportation in the two spaces may be different. Here we rescale the two datasets to make the actions in $\mathcal X$ and $\mathcal Y$ comparable (see \hyperref[sec:method-data-normalization]{Data normalization}), and then assume equal trade-off weights $\alpha$ and $\beta$ in two spaces.

\subsection*{Trajectory-synchronized UnSyncOT vs Density-synchronized UnSyncOT}

Note that in our original formulation of UnSyncOT \cite{cang2026unsyncot}, the feasibility condition reads
\begin{align}\label{eqn:feasibility2}
\left\{
\begin{array}{@{}l@{}}
\partial_t\rho_t+\alpha\nabla_x\cdot(\rho_tu_t)
=\beta\rho_tg_t, \ \rho_{t_k} = \widehat{\rho}_{k}, \ u_t\cdot n_x = 0 \text{\ on }\partial\mathcal X;\\[2pt]
\partial_t\xi_t+\alpha\nabla_y\cdot(\xi_tv_t)
=\beta\xi_th_t, \ \ \xi_{t_k} = \widehat{\xi}_{k}, \ v_t\cdot n_y = 0 \text{\ on }\partial\mathcal Y;\\[2pt]
\xi_t=(\mathbf{T}_t)_{\#}\rho_t,\quad
h_t(\mathbf T_t(x))=g_t(x),
\end{array}
\right.
\end{align}
without the velocity constraint
\begin{align}\label{eqn:velocityconstraint}
v_t(\mathbf{T}_t(x)) = \nabla_x\mathbf{T}_t(x)u_t(x) + \alpha^{-1}\partial_t \mathbf{T}_t(x).
\end{align}
However condition (\ref{eqn:feasibility2}) does not automatically imply the velocity constraint (\ref{eqn:velocityconstraint}). Indeed, given condition (\ref{eqn:feasibility2}), there may be other choices on the relation between the primary and secondary velocities as below such that the secondary continuity equation still holds,
\begin{align}\label{eqn:velocityconstraint2}
v_t(\mathbf{T}_t(x)) = \nabla_x\mathbf{T}_t(x)u_t(x) + \alpha^{-1}\partial_t \mathbf{T}_t(x) + z_t(\mathbf{T}_t(x)),
\end{align}
for any $z_t$ such that $\nabla_y\cdot(\xi_t z_t)=0$. When minimizing the coupled energy across the two spaces, we need to take $z_t$ such that the corresponding $v_t$ gives the smallest possible kinetic energy. To this end, the secondary velocity $v_t$, according to Otto's theory \cite{villani2009optimal, figalli2021invitation}, should be taken in the following form:
\begin{align*}
v_t^{\mathrm{energy}}(y) = \nabla \Phi(y), \quad
\mathrm{with\ }
-\alpha\nabla\cdot(\xi_t\nabla\Phi) = \partial_t\xi_t - \beta \xi_t h_t.
\end{align*}
We call $v_t^{\mathrm{energy}}(y)$ the energy-minimizing velocity. In contrast, we call equation (\ref{eqn:velocityconstraint}) the map-induced velocity. Correspondingly, we distinguish two formulations of the UnSyncOT. The one in (\ref{eqn:UnSyncOT01}) is called trajectory-synchronized UnSyncOT, and the other formulation with the feasibility condition (\ref{eqn:feasibility2}) is called density-synchronized UnSyncOT.

The difference between trajectory-synchronized and density-synchronized UnSyncOT can be illustrated by an example as follows. Consider $\rho_t$ as a fixed standard Gaussian $\rho_t = \mathcal{N}(\cdot | 0, I)$, and $\mathbf{T}_t$ being a uniform rotation transform. Then the induced $\xi_t = (\mathbf{T}_t)_{\sharp}\rho_t$ becomes a uniformly rotating Gaussian which is still $\mathcal{N}(\cdot | 0, I)$. Therefore, density-synchronized UnSyncOT gives the optimizer $\rho_t = \mathcal{N}(\cdot | 0, I), u_t = 0, g_t = 0$. But the trajectory-synchronized UnSyncOT gives a nontrivial solution as the induced secondary trajectories are circular paths.

Since this work focuses on the trajectory synchronization across the RNA and ATAC spaces, we adopt the formulation of trajectory-synchronized UnSyncOT (\ref{eqn:UnSyncOT01}). For simplicity, we continue to refer to this formulation as UnSyncOT throughout the paper. A detailed comparison of the two UnSyncOT formulations can be found in \cite{cang2026unsyncot}.

\subsection*{Reformulation of UnSyncOT as a Single-space OT}

Since the two molecular spaces are linked by the feasibility conditions in (\ref{eqn:UnSyncOT01}), UnSyncOT can be reformulated as a single-space dynamical OT problem \cite{cang2026unsyncot}, and then further approximated by the continuous normalizing flow (CNF) model \cite{tong2020trajectorynet}. We summarize the reformulations as follows.
\begin{theorem}\label{theorem:UnSyncOT_reformulation}
    UnSyncOT (\ref{eqn:UnSyncOT01}) can be reformulated as a single-space dynamical OT
    \begin{equation}\label{single_space_ot}
    \begin{aligned}
    &\inf_{(\rho_t,u_t, g_t)}
    \frac{1}{2}\int_0^T 
    \left[
    \int_{\mathcal X}
    c_x\alpha\ \|u_t\|^2 +
    c_y\alpha\|(\nabla_x\mathbf{T}_t)u_t + \alpha^{-1}\partial_t\mathbf{T}_t\|^2 +
    \beta g_t^2\right)\rho_t\,dx dt
    \\[4pt]
    \text{subject to}\quad
    &\partial_t\rho_t+\alpha\nabla_x\cdot(\rho_tu_t)
    =\beta\rho_tg_t, \ \rho_{t_k} = \widehat{\rho}_{k},\ k=0,\cdots,K, \ u_t\cdot n_x = 0 \text{\ on }\partial\mathcal X.
    \end{aligned}
    \end{equation}
    Let $M_0 = \widehat{\rho}_0(\mathcal{X})$. Formulation (\ref{single_space_ot}) can be approximated by CNF as
    \begin{align}
        \inf_{(u_t,g_t)} \mathbb{E}_{X_0\sim M_0^{-1}\widehat{\rho}_0}\left[ \frac{M_0}{2}\int_0^T \Big( c_x\alpha\|u_t(X_t)\|^2 + c_y\alpha \|v_t(Y_t)\|^2 + \beta g_t(X_t)^2 \Big)w_t(X_0) dt \right] \notag\\
        + \lambda \sum_{i=1}^N \mathbb{D}\left(\frac{\rho_{t_i}}{\rho_{t_i}(\mathcal{X})},\frac{\widehat{\rho}_i}{\widehat{\rho}_i(\mathcal{X})}\right), \label{eqn:CNF}
    \end{align}
    in which
    \begin{align}
      &\dot{X}_t(x)
      =\alpha u_t(X_t(x)), \quad X_0(x) = x \sim M_0^{-1}\widehat{\rho}_0; \label{eqn:flow_primary}\\
      & Y_t = \mathbf{T}_t(X_t)
      \mathrm{\ such \ that\ }
      \dot{Y}_t
      =\alpha v_t(Y_t) = \alpha[\nabla\mathbf{T}_t u_t + \alpha^{-1}\partial_t \mathbf{T}_t]; \label{eqn:flow_secondary}
      \\
      &\dot{(\log w_t(x))}
      =\beta g_t(X_t(x)), \ w_0(x) \equiv 1 
      \mathrm{\ such \ that\ }
      w_t(x) = \exp\left(\beta \int_0^t g_s(X_s(x))ds\right). \label{eqn:relativemass}
    \end{align}
    Here $X_t(x)$ is a primary RNA trajectory determined by the flow equation (\ref{eqn:flow_primary}); $Y_t$ is the associated secondary ATAC trajectory induced by $X_t$ through $\mathbf{T}_t$, whose velocity is obtained by chain rule; $w_t(x)$ is the relative mass along the trajectory $X_t(x)$. $\mathbb{D}$ is a discrepancy measure between predicted and observed data to approximately fulfill the marginal constraints $\rho_{t_k} = \widehat{\rho}_k$ for $k=1,\cdots,K$. 
\end{theorem}
    \begin{proof}
        The single-space formulation (\ref{single_space_ot}) follows Theorem 3.1 of \cite{cang2026unsyncot}. The CNF approximation follows Theorem 4.1 of \cite{tong2020trajectorynet} and Theorem of \cite{sha2024reconstructing}. 
    \end{proof}

    In what follows, UnSyncOT (\ref{single_space_ot}) will be solved by the CNF approximation (\ref{eqn:CNF}) together with the flow equations (\ref{eqn:flow_primary})-(\ref{eqn:relativemass}). To this end, we introduce the neural network approximations for $(u_t, g_t)$ and the cross-modality map $\mathbf{T}_t$. In the training loss defined later in (\ref{eqn:loss_Lk}), we take $\mathbb{D}$ to be the Sinkhorn divergence, see equations (\ref{eqn:loss_marginal_X})-(\ref{eqn:loss_marginal_Y}).

\subsection*{Neural Network Approximations of UnSyncOT Variables}
\label{sec:method-vector-field}
Since the CNF minimization (\ref{eqn:CNF}) is over $(u_t, g_t)$, we parameterize them by fully connected neural networks,
\[u^\theta_t(x):\mathcal X\times[0,T]\to\mathbf R^{d_x}, \quad g^\phi_t(x):\mathcal X\times[0,T]\to\mathbf R.
\]
We sample $B$ initial RNA cells $X_0^{(i)}\sim M_0^{-1}\widehat{\rho}_0$ with initial mass $w_0^{(i)} = 1/B$. Then the trajectories $\{(X_t^{(i)}, w_t^{(i)})\}_{i=1}^B$ are obtained by the flow equations
\begin{align}
  \dot{X}_t^{(i)}
  &=\alpha u^\theta_t(X_t^{(i)}), \quad X_0^{(i)}\sim M_0^{-1}\rho_0;\label{eqn:flow_position}\\
  \dot{w}_t^{(i)} 
  &=\beta g_t^\phi(X_t^{(i)})w_t^{(i)}, \quad w_0^{(i)}= 1/B. \label{eqn:flow_mass}
\end{align}
With a slight abuse of notation, $w_t$ in (\ref{eqn:relativemass}) denotes the relative mass along trajectory $X_t$, whereas $w_t^{(i)}$ in (\ref{eqn:flow_mass}) denotes the actual mass carried by the $i$-th trajectory $X_t^{(i)}$.

The map $\mathbf{T}_t$ between RNA and ATAC spaces is approximated by a pretrained and fixed neural network $\mathbf{T}_t^{\eta}$ such that the secondary trajectories are taken as $Y_t^{(i)} = \mathbf T^\eta_t(X_t^{(i)})$. ATAC cell $Y_t^{(i)}$ carries the same mass $w_t^{(i)}$ as that of the corresponding RNA cell $X_t^{(i)}$.
We calculate kinetic energy in the secondary space by finite differences
(see \hyperref[sec:method-loss]{Training loss}).
We also compared the alternative mapping architectures (see
\hyperref[sec:method-cross-modal-map]{Cross-modality map}).

\subsection*{Training loss}
\label{sec:method-loss}

The total training loss is $\mathcal{L}(\theta,\phi,\eta) = \sum_{k=0}^{K-1} \mathcal{L}_k(\theta,\phi,\eta)$, with $\mathcal{L}_k$ being the total training loss over interval $[t_k, t_{k+1}]$. We define $\mathcal{L}_k$ as a sum of several losses as below.

For each training iteration, we sample $B$ RNA cells $\{X_0^{(i)}\}_{i=1}^B$ from the observed RNA distribution $\widehat{\rho}_0$ at $t_0$ and initialize their masses as $w_0^{(i)}=1/B$. 
Observed cells at intermediate stages are used only to evaluate the losses.

\medskip
\noindent\textbf{Kinetic energy loss.}
Define $E^{(i)}_{k,\mathcal{X}}(t)= \int_{t_k}^t \frac{\alpha}{2}\|u^{\theta}_t(X^{(i)}_t)\|^2 w^{(i)}_t dt$ over $[t_k,t_{k+1}]$. The primary kinetic energy in CNF formulation (\ref{eqn:CNF}) driven by the flow (\ref{eqn:flow_position})-(\ref{eqn:flow_mass}) is given as
\[
\mathcal{L}^{\mathcal{X}}_{\mathrm{kinetic}}(k) = \sum_{i=1}^B E^{(i)}_{k,\mathcal{X}}(t_{k+1}). 
\]

In the secondary space, we know that $Y_t^{(i)} = \mathbf{T}_t^{\eta}(X_t^{(i)})$. Note from (\ref{eqn:flow_secondary}) that 
\[
\dot{Y}_t^{(i)} = \alpha\nabla\mathbf{T}_t^{\eta}(X_t^{(i)}) u_t^{\theta}(X_t^{(i)}) + \partial_t \mathbf{T}_t^{\eta}(X_t^{(i)}) = \frac{d}{d\epsilon} T^{\eta}_{t+\epsilon}(X_t^{(i)}+\epsilon\alpha u_t^{\theta}(X_t^{(i)})) \Big
|_{\epsilon = 0},
\]
we can therefore approximate the secondary velocity by finite difference and obtain the secondary kinetic energy over $[t_k,t_{k+1}]$ as follows
\begin{align}\label{eqn: loss_kinetic_Y}
\mathcal{L}^{\mathcal{Y}}_{\mathrm{kinetic}}(k) =\sum_{i=1}^{B}
\int_{t_k}^{t_{k+1}} \frac{\alpha}{2}\|v_t^{(i)}\|^2 w_t^{(i)} dt 
\approx \sum_{i=1}^B \sum_{l=0}^{L-1} w_{s_l}^{(i)} \frac{\|Y_{s_{l+1}}^{(i)} - Y_{s_{l}}^{(i)}\|^2}{2\alpha(s_{l+1}-s_l)},
\end{align}
in which $t_k=s_0<s_1<\cdots<s_L=t_{k+1}$ is a subpartition over $[t_k, t_{k+1}]$. For simplicity, we assume  the total number of subintervals $L$ is independent of $k$.

\medskip
\noindent\textbf{Growth loss.}
Defining $E^{(i)}_{k,\mathrm{growth}}(t)= \int_{t_k}^t \frac{\beta}{2}|g^{\phi}_t(X^{(i)}_t)|^2 w^{(i)}_t dt$ over $[t_k,t_{k+1}]$, the growth loss in CNF formulation (\ref{eqn:CNF}) driven by the flow (\ref{eqn:flow_position})-(\ref{eqn:flow_mass}) is given as
\[
\mathcal{L}_{\mathrm{growth}}(k) = \sum_{i=1}^B E^{(i)}_{k,\mathrm{growth}}(t_{k+1}). 
\]
Note that the growth relation $h_t(\mathbf T_t(x))=g_t(x)$ guarantees that the primary and secondary growth losses are identical. Therefore, the two growth losses sum into one term as shown in equation (\ref{eqn:CNF}).

\medskip
\noindent\textbf{Reference marginal matching.}
For the marginal constraints $\rho_{t_k} = \widehat{\rho}_k$, $\xi_{t_k} = \widehat{\xi}_k$ for $k=0,\cdots, K$, we measure the discrepancy between the predicted distribution $\{X_{t_{k+1}}^{(i)}\}_{i=1}^{B}$ and observed distributions $\widehat{\rho}_{k+1}$ using Sinkhorn divergences. Here Sinkhorn divergence is defined as
\[
S_{\epsilon}(\mu,\nu) = \mathrm{OT}_{\epsilon}(\mu,\nu) - 
\frac{1}{2}\mathrm{OT}_{\epsilon}(\mu,\mu) -
\frac{1}{2}\mathrm{OT}_{\epsilon}(\nu,\nu)
\]
in which $\mathrm{OT}_{\epsilon}$ denotes the standard entropy-regularized Kantorovich optimal transport cost, with $\epsilon$ being the entropy regularization constant.

Let $\mathcal B_{k+1}^{\mathcal{X}}=\{\widehat{X}_{t_{k+1}}^{(j)}\}_{j=1}^{m_{k+1}}$ and $\mathcal B_{k+1}^{\mathcal{Y}}=\{\widehat{Y}_{t_{k+1}}^{(j)}\}_{j=1}^{n_{k+1}}$ denote mini-batches from the observed RNA data $\widehat{\rho}_{k+1}$ and ATAC data $\widehat{\xi}_{k+1}$. After normalizing the predicted cell weights as
$
\bar w_t^{(i)}=\frac{w_t^{(i)}}{\sum_{q=1}^{B}w_t^{(q)}},
$
we define the reference marginal matching loss as
\begin{align}
    &\mathcal{L}_{\mathrm{ref}}^{\mathcal{X}}(k) = 
    S_{\epsilon_{k+1}^{\mathcal{X}}}\left(\sum_{i=1}^B \bar{w}_{t_{k+1}}^{(i)}\delta_{X_{t_{k+1}}^{(i)}}, \sum_{j=1}^{m_{k+1}} \frac{1}{m_{k+1}}\delta_{\widehat{X}_{t_{k+1}}^{(j)}} \right); \label{eqn:loss_marginal_X}\\
    &\mathcal{L}_{\mathrm{ref}}^{\mathcal{Y}}(k) = 
    S_{\epsilon_{k+1}^{\mathcal{Y}}}\left(\sum_{i=1}^B \bar{w}_{t_{k+1}}^{(i)}\delta_{Y_{t_{k+1}}^{(i)}}, \sum_{j=1}^{n_{k+1}} \frac{1}{n_{k+1}}\delta_{\widehat{Y}_{t_{k+1}}^{(j)}} \right).\label{eqn:loss_marginal_Y}
\end{align}
Here we use half the squared Euclidean distance between $X_{t_{k+1}}^{(i)}$ and $\widehat{X}_{k+1}^{(j)}$ for the cost matrix in $\mathrm{OT}_{\epsilon}$ in $\mathcal{X}$. Similarly for that in $\mathcal{Y}$. The entropy regularization scales $\epsilon_k^{\mathcal{X},\mathcal{Y}}$ are given by the rule in
\hyperref[sec:method-sinkhorn-scale-preprocessing]{Sinkhorn scale preprocessing}.

\medskip
\noindent\textbf{Unbalanced mass matching.}
Let $L_k$ denote the observed population size at $\widehat{\rho}_k$,  and define
$
r_{k}=L_{k}/L_0.
$
Here $L_k$ is the estimated number of cells for a specific system at time $t_k$. If $L_k$ is known from published work, we use it to define the population size ratio $r_k$; otherwise, we set $r_k = 1$. For the five datasets we studied in this paper, mouse gastrulation and human cerebral organoids have population-size priors, which define nontrivial $r_k$; scMultiSim, Mouse secondary palate, and mouse pancreas do not have the population-size prior, so we take $r_k = 1$. See \hyperref[sec:method-data-preprocessing]{Data preprocessing} for the details.

Following the idea in DeepRUOT \cite{zhang2024learning}, we assign each observed RNA cell $\widehat{X}_{k+1}^{(j)}$ to its nearest predicted RNA cell $X_{t_{k+1}}^{(i)}$: $\pi_j= \operatorname*{arg\,min}_{1\leq i\leq B} \|\widehat{X}_{k+1}^{(j)} - X_{t_{k+1}}^{(i)}\|^2.$ For each predicted RNA cell $X_{t_{k+1}}^{(i)}$, we count the observed RNA cells $\widehat{X}_{k+1}^{(j)}$ assigned to it and divide this count by the total number \(m_{k+1}\) of cells in the reference batch $\mathcal{B}^{\mathcal{X}}_{k+1}$. Multiplying the resulting proportion by \(r_{k+1}\) gives the target mass $\widehat w_{k+1}^{(i)}$ of that predicted RNA cells $X_{t_{k+1}}^{(i)}$,
\begin{equation*}
  \widehat w^{(i)}_{k+1}=
  \frac{r_{k+1}}{m_{k+1}}
  \sum_{j=1}^{m_{k+1}}\mathbf 1_{\{\pi_j=i\}}.
\end{equation*}
In other words, we use the predicted RNA cells $\{X_{t_{k+1}}^{(i)}\}_{i=1}^B$ as the generating points to construct a Voronoi tessellation of the RNA space. The target mass $\widehat w_{k+1}^{(i)}$ assigned to each generating point $X_{t_{k+1}}^{(i)}$ is then defined as \(r_{k+1}\) times the number of observed RNA cells in the Voronoi region of $X_{t_{k+1}}^{(i)}$, divided by the total number \(m_{k+1}\) of observed RNA cells. Hence we have $\sum \widehat w_{k+1}^{(i)} = r_{k+1}$.

We penalize the difference between the predicted and target masses:
\begin{equation}\label{eqn:loss_mass}
  \mathcal L_{\mathrm{mass}}(k) =
  \sum_{i=1}^{B}
  \left(w_{t_{k+1}}^{(i)}- \widehat w_{k+1}^{(i)}\right)^2.
\end{equation}
 when $\mathcal L_{\mathrm{mass}}(k)=0$, $\sum_{i=1}^{B}w_{t_{k+1}}^{(i)} = \sum_{i=1}^{B}\widehat w_i^{k+1} = r_{k+1}$, so the predicted total mass matches the observed mass ratio. 
 
 \medskip
\noindent\textbf{Manifold constraint loss.}
To keep the trajectories $X_{t}^{(i)}$ and $Y_{t}^{(i)}$ within the support of the observed data manifold, we penalize predicted trajectories when they become far from observed cells. To this end, we define observed data manifold over $[t_k,t_{k+1}]$ as
\[
\mathcal S_k^\mathcal{X}=\mathcal B_k^\mathcal{X}\cup\mathcal B_{k+1}^\mathcal{X},
\quad
\mathcal S_k^\mathcal{Y}=\mathcal B_k^\mathcal{Y}\cup\mathcal B_{k+1}^\mathcal{Y}.
\]
Let $d_{K_{\mathrm{nn}}}(z,\mathcal S)$ denote the mean of the Euclidean distances from $z$ to its $K_{\mathrm{nn}}$ nearest neighbors in $\mathcal S$. We introduce a penalty when this distance from trajectories to the data manifolds $S_k^\mathcal{X}$ and $S_k^\mathcal{Y}$ exceeds
a threshold $\delta\geq0$:
\begin{equation}\label{eqn:loss_manifold}
\begin{aligned}
\mathcal L_{\mathrm{manifold}}^\mathcal{X}(k)
&=
\frac{1}{B(L+1)}
\sum_{i=1}^{B}\sum_{l=0}^{L}
\left[
d_{K_{\mathrm{nn}}}(X_{s_l}^{(i)},\mathcal S_k^\mathcal{X})-\delta
\right]_+,\\
\mathcal L_{\mathrm{manifold}}^\mathcal{Y}(k)
&=
\frac{1}{B(L+1)}
\sum_{i=1}^{B}\sum_{l=0}^{L}
\left[
d_{K_{\mathrm{nn}}}(Y_{s_l}^{(i)},\mathcal S_k^\mathcal{Y})-\delta
\right]_+,
\end{aligned}
\end{equation}
where $[a]_+=\max(a,0)$ and $t_k = s_0 < \cdots<s_l<\cdots<s_L = t_{k+1}$ is a subpartition over $[t_k, t_{k+1}]$, the same as we use to define $\mathcal{L}^{\mathcal{Y}}_{\mathrm{kinetic}}(k)$ in (\ref{eqn: loss_kinetic_Y}). Both losses in (\ref{eqn:loss_manifold}) are averaged uniformly over the cells, including in the unbalanced model.

\medskip
\noindent\textbf{Total loss.}
Summing all the losses defined above, the total loss is
\begin{align}\label{eqn:loss_Lk}
\mathcal{L}(\theta,\phi,\eta) = \sum_{k=0}^{K-1}\mathcal L_k
={} \sum_{k=0}^{K-1} \Big( &c_x\left[
\lambda_E\mathcal L_{\mathrm{kinetic}}^\mathcal{X}(k)
+\lambda_D\mathcal L_{\mathrm{manifold}}^\mathcal{X}(k)
\right] \notag\\
&+c_y\left[
\lambda_E\mathcal L_{\mathrm{kinetic}}^\mathcal{Y}(k)
+\lambda_D\mathcal L_{\mathrm{manifold}}^\mathcal{Y}(k)
\right] \notag\\
&+\lambda_x\mathcal L_{\mathrm{ref}}^\mathcal{X}(k)
+\lambda_y\mathcal L_{\mathrm{ref}}^\mathcal{Y}(k)
+\lambda_G\mathcal L_{\mathrm{growth}}(k)
+\lambda_M\mathcal L_{\mathrm{mass}}(k) \Big).
\end{align}
Here, $\lambda_E$, $\lambda_D$ and $\lambda_M$ weight the kinetic energy, manifold and mass-matching losses, respectively, $\lambda_x$ and $\lambda_y$ weight RNA and ATAC reference marginal matching. They are all independent of $k$.

Throughout the paper, we take $\alpha = 1$ for the trade-off between transportation and teleportation in UnSyncOT (\ref{eqn:UnSyncOT01}).  For the balanced case, we set $\beta=0$ and therefore discard the growth and mass-matching terms. The weight $c_y$ controls the kinetic energy and manifold losses in ATAC space. Note that when taking $c_y=0$ in the COATI loss (\ref{eqn:loss_Lk}), the reference marginal constraint $\mathcal{L}^{\mathcal{Y}}_{\mathrm{ref}}$ in ATAC space is still kept. Therefore this extreme case does not recover the unbalanced dynamical OT only in RNA space, which we refer to as the RNA-only model. The RNA-only ablation requires removing all ATAC loss terms in $\mathcal{L}_k$ (\ref{eqn:loss_Lk}).

\subsection*{Adaptive reference marginal matching weights $\lambda_x$ and $\lambda_y$}
\label{sec:method-adaptive-marginal-weight}

UnSyncOT (\ref{eqn:UnSyncOT01}) imposes exact marginal constraints, which we replace with reference marginal matching penalties $\mathcal{L}_{\mathrm{ref}}^{\mathcal{X},\mathcal{Y}}$ in (\ref{eqn:loss_Lk}) during training. However, choosing the weights $\lambda_{x,y}$ for these losses is tricky. If the weights are too small, the predicted distributions
may not adequately match the observed $\widehat \rho_k$ and $\widehat \xi_k$. If the weights are too large, marginal matching can dominate training, limiting the effective optimization of the kinetic energy and other loss terms. We therefore tune the weights according to whether $\mathcal{L}_{\mathrm{ref}}^{\mathcal{X},\mathcal{Y}}(k)$ exceed certain given tolerances $\tau_{k+1}^{\mathcal{X},\mathcal{Y}}$. 

More specifically, we compute the tolerances \(\tau_k^{\mathcal X}\) and \(\tau_k^{\mathcal Y}\) separately at each observation time \(t_k\). For RNA, we draw \(2Q\) independent mini-batches \(\{\widehat X_{k,q}^{(i)}\}_{i=1}^{m_k}\), \(q=1,\ldots,2Q\), from the observed snapshot \(\widehat\rho_k\). We group these mini-batches into \(Q\) pairs and define the tolerance as their mean Sinkhorn divergence:
\begin{align}\label{eqn:tolerance}
\tau_k^{\mathcal X}
=
\frac{1}{Q}\sum_{q=1}^{Q}
S_{\epsilon_k^{\mathcal X}}
\left(
\frac{1}{m_k}\sum_{i=1}^{m_k}
\delta_{\widehat X_{k,2q-1}^{(i)}},
\frac{1}{m_k}\sum_{i=1}^{m_k}
\delta_{\widehat X_{k,2q}^{(i)}}
\right),
\end{align}
where \(\epsilon_k^{\mathcal X}\) is specified in \hyperref[sec:method-sinkhorn-scale-preprocessing]{Sinkhorn scale preprocessing}. This tolerance estimates the average discrepancy arising from independently sampling two mini-batches from the same observed snapshot. It therefore provides an empirical reference scale for the mini-batch sampling variability. We use \(\tau_k^{\mathcal X}\) as the tolerance when comparing the predicted RNA cells \(\{X_{t_k}^{(i)}\}_{i=1}^{B}\) with an observed mini-batch \(\mathcal B_k^{\mathcal X} = \{\widehat X_{k}^{(i)}\}_{i=1}^{m_k}\) so that the resulting discrepancy between them is comparable to the average discrepancy between two mini-batches drawn from the observed data $\widehat{\rho}_k$ itself. The ATAC tolerance \(\tau_k^{\mathcal Y}\) is computed analogously using the observed ATAC snapshot and the regularization parameter \(\epsilon_k^{\mathcal Y}\).

At the $n$-th training step, take $\Delta^{\mathcal{X}}:=\max_{0\le k\le K-1}(\mathcal{L}^{\mathcal{X}}_{\mathrm{ref}}(k;\theta^n,\phi^n) - \tau^{\mathcal{X}}_{k+1})$ , then we update the reference marginal matching weight $\lambda_x$ in RNA space by
\begin{equation}
\lambda_x^{n+1}
\leftarrow
\operatorname{clip}
\left(
\lambda_x^n + h_\lambda\Delta^{\mathcal{X}};
\lambda_{\min}^{\mathcal{X}},
\lambda_{\max}^{\mathcal{X}}
\right).
\end{equation}
Here, $h_\lambda>0$ is the step size for the update of $\lambda_x$. In all the experiments, we set $h_\lambda = 100$. The parameters $\lambda_{\min}^\mathcal{X}$ and $\lambda_{\max}^\mathcal{X}$ specify the lower and upper bounds on $\lambda_x$, respectively. We typically set the lower and upper bounds to 10 and 100, respectively, for both modalities. In mouse palate and gastrulation experiments, we reduced the RNA bounds to 5 and 50 to place greater relative emphasis on ATAC distributional matching. When $\Delta^{\mathcal{X}}>0$, we increase $\lambda_x$ at the next training step, up to $\lambda_{\max}^{\mathcal X}$, to give greater weight to RNA marginal matching. When $\Delta^{\mathcal{X}}<0$, we decrease $\lambda_x$ but not below $\lambda_{\min}^{\mathcal{X}}$.
An analogous update rule is applied to $\lambda_y$ for the ATAC space.

\subsection*{Sinkhorn scale preprocessing}
\label{sec:method-sinkhorn-scale-preprocessing}

The Sinkhorn regularization scale was selected separately for each modality and observed stage, using the normalized coordinates (see \hyperref[sec:method-data-normalization]{Data normalization}). At each time \(t_k\), we randomly subsampled observed RNA and ATAC cells and calculated the median pairwise Euclidean distance within each subsample:
\[
\begin{aligned}
c_k^{\mathcal X}
=
\operatorname{median}_{\substack{i,j\in\mathcal I_k^{\mathcal X}\\i<j}}
\left\|\widehat X_k^{(i)}-\widehat X_k^{(j)}\right\|_2, \quad
c_k^{\mathcal Y}
=
\operatorname{median}_{\substack{i,j\in\mathcal I_k^{\mathcal Y}\\i<j}}
\left\|\widehat Y_k^{(i)}-\widehat Y_k^{(j)}\right\|_2,
\end{aligned}
\]
where \(\mathcal I_k^{\mathcal X}\) and \(\mathcal I_k^{\mathcal Y}\) denote the subsampled cell indices. We then set
\[
\varepsilon_k^b=c_{\varepsilon}c_k^b,
\qquad
c_{\varepsilon}=0.05,
\qquad
b\in\{\mathcal X,\mathcal Y\}.
\]
These parameters were computed once after coordinate normalization and remain fixed throughout training. For each stage and modality, the same scale was used for the reference marginal matching loss in (\ref{eqn:loss_marginal_X})-(\ref{eqn:loss_marginal_Y}) and for sampling-based tolerance in (\ref{eqn:tolerance}), ensuring that both quantities were calculated using the same Sinkhorn divergence.

\subsection*{Data preprocessing}
\label{sec:method-data-preprocessing}

COATI does not require explicit pairing between RNA and ATAC cells. In our training pipeline, paired observations are used to learn the cross-modality map $\mathbf T^{\eta}_t$. The scMultiSim, mouse gastrulation, mouse pancreas and mouse secondary palate datasets contain paired RNA and ATAC profiles. For the unpaired human cerebral organoid data, we train the cross-modal map $\mathbf T^{\eta}_t$ using computationally paired metacells.

\medskip
\noindent\textbf{scMultiSim.}
We used 10,000 paired RNA and ATAC profiles simulated with scMultiSim \cite{scmultisim}. Cells were grouped by pseudotime into four snapshots containing 2,006, 2,001, 3,008 and 2,985 cells at $t=0,1,2,3$, respectively. The empirical distribution at each snapshot was normalized to unit mass.

RNA counts were transformed using \(\log_2(1+x)\). Each gene was then standardized across all cells before PCA, and the first 10 principal components were used as the RNA representation. ATAC features were standardized across all cells. We calculated the correlation between sequencing depth and the cell scores on each principal component. The second component was excluded because it was strongly associated with sequencing depth. Components 1 and 3–9 were used as a reduced dimensional representation.

\medskip
\noindent\textbf{Mouse gastrulation.}
For the gastrulation multiome analysis, we obtained the RNA, ATAC peak and ATAC TSS count matrices from the Bioconductor \texttt{MouseGastrulationData} package using \texttt{RAMultiomeData} \cite{griffiths2026mousegastrulationdata}. Cell annotations followed the nomenclature of the mouse gastrulation and early organogenesis atlas \cite{pijuanSala2019single}. We exported the matrices and metadata from the \texttt{SingleCellExperiment} objects and converted them to AnnData format. CRISPR perturbation samples were excluded from training and reserved for perturbation analysis. We assigned model times $\{0.0, 1.0, 2.0, 2.5\}$ to $\{\mathrm{E}7.5, \mathrm{E}8.0, \mathrm{E}8.5, \mathrm{E}8.75\}$, respectively. For unbalanced COATI, we constructed the population-size prior by using published estimates of the total number of cells per embryo, approximately 15,000 cells at E7.5, 60,000 at E8.0, and 90,000 at E8.5 \cite{qiu2022systematic, klein2025mapping}.
For E8.75, we assumed a population mass 11 times that at E7.5, corresponding to 165,000 cells on this reference scale. Thus the resulting
relative population masses used by COATI were
\begin{align}\label{eqn:population_size_prior}
L_{\mathrm{E7.5}}:L_{\mathrm{E8.0}}:L_{\mathrm{E8.5}}:L_{\mathrm{E8.75}}
=1:4:6:11,
\end{align}
which were used in the mass matching loss (\ref{eqn:loss_mass}).

For RNA preprocessing, we selected 3,000 highly variable genes from raw counts using the Seurat-v3 procedure \cite{stuart2019comprehensive}. Counts were normalized to a total of \(10^4\) per cell and log-transformed. We then applied PCA to the expression matrix of the selected genes and used the first 50 principal components. The dataset contained 9,018, 7,796, 11,470 and 7,219 measured cell profiles at E7.5, E8.0, E8.5 and E8.75, respectively. Note that these sample sizes were only used to construct the empirical distributions and were not interpreted as total embryo population sizes. For unbalanced COATI, we separately specified a population-size prior (\ref{eqn:population_size_prior}) for the mass matching.

For the matched ATAC data, we kept peaks detected in more than $1\%$ and less than $50\%$ of cells, leaving 181,039 of the original 192,251 peaks. We computed term-frequency inverse-document-frequency (TF-IDF) features and selected the 25,000 peaks with the highest variance across cells. Truncated singular value decomposition (SVD) yielded 50 latent semantic indexing (LSI) components. We calculated the absolute Spearman correlation between the cell scores on each component and per-cell ATAC depth. The first component was excluded because it was strongly correlated with depth. Components 2-15 were used as a 14-dimensional ATAC representation.

\medskip
\noindent\textbf{Mouse pancreas.}
For the mouse pancreas analysis, we used the processed moscot dataset \cite{klein2025mapping}. The E14.5, E15.5 and E16.5 snapshots contained 9,029, 10,333 and 3,242 measured cell profiles, respectively. We assigned model times \(\{0.0, 1.0, 2.0\}\) to $\{ \mathrm{E}14.5, \mathrm{E}15.5, \mathrm{E}16.5\}$, respectively.

For RNA, we used the 50-dimensional PCA coordinates provided in the processed RNA object as the representation. For ATAC, we used the 22-dimensional PoissonVI representation stored under \(X_{\text{poissonvi}}\) in the released object. This representation was generated using the scvi-tools framework \cite{gayoso2022python}. We used the released coordinates directly, without retraining the model.

\medskip
\noindent\textbf{Mouse secondary palate.}
For the palate multiome analysis \cite{yan2024single}, we assigned model times \(\{0.0, 1.0, 1.5, 2.0\}\) to $\{ \mathrm{E}12.5, \mathrm{E}13.5, \mathrm{E}14.0, \mathrm{E}14.5\}$, respectively. The four snapshots contained 2,570, 5,692, 6,638 and 4,933 measured cell profiles, respectively.

For RNA preprocessing, we scaled the expression matrix of the selected highly variable genes and clipped the scaled values to the range \([-10,10]\). We then applied PCA and kept the first 40 principal components as the RNA representation.

For the matched ATAC data, we kept peaks detected in at least $1\%$ of cells, leaving 84,395 of the original 123,807 peaks. Peak counts were binarized and transformed using TF-IDF with sublinear term-frequency scaling and \(l_2\) normalization. We then applied truncated SVD to obtain LSI components. The first component was excluded because it was strongly correlated with sequencing depth. Components 2–16 were retained, yielding a 15-dimensional ATAC representation.

\medskip
\noindent\textbf{Human cerebral organoids.}
For the human cerebral organoid analysis, we used the developmental scRNA-seq and scATAC-seq time course of Fleck et al. \cite{fleck2023inferring}, which included three induced pluripotent stem cell lines and one embryonic stem cell line. RNA and ATAC were measured in different cells from the same organoid suspensions. We therefore used the computationally paired RNA–ATAC metacells released by Fleck et al. \cite{fleck2023inferring}. These metacells were constructed by integrating RNA expression and ATAC-derived gene activity within each combination of developmental stage and stem cell line, followed by bipartite matching in the shared space.

We selected developmental days $(4, 7, 9, 11, 12, 18, 21)$ and assigned model times according to \(t=(d-4)/10\). The corresponding model times were
$
(0.0,\,0.3,\,0.5,\,0.7,\,0.8,\,1.4,\,1.7).
$
The seven snapshots contained 501, 1,309, 1,630, 1,560, 4,100, 5,842 and 5,721 paired metacells, respectively. Day 16 was excluded because it contained few metacells and showed an obvious discontinuity from neighboring stages in both modalities.

For the unbalanced models, we separately specified a population-size prior using measurements of the total number of cells per organoid from Lindenhofer et al. \cite{lindenhofer2024cerebral}. We averaged the biological replicates at each measured time point, linearly interpolated the measurements to the selected developmental days and normalized them by the value in day 4. This yielded relative population masses of
$
(1.00,\,2.92,\,4.15,\,5.36,\,5.96,\,22.94,\,33.99).
$
These values provided a prior for population growth, whereas the metacell counts specified the number of observations available at each stage.

For RNA preprocessing, highly variable genes were selected, and their expression values were scaled and corrected for technical factors, cell cycle, aggregation size and stem cell line. We then applied PCA and kept the first 30 principal components as the RNA representation.

For ATAC preprocessing, metacell peak accessibility was binarized and transformed using TF-IDF, followed by truncated SVD to obtain LSI components. Components dominated by sequencing depth or showing low signal were excluded, leaving 12 components. The resulting coordinates were further corrected for technical factors, aggregation size and stem cell line.

\subsection*{Data normalization}
\label{sec:method-data-normalization}
The reduced RNA and ATAC embeddings have different dimensions and numerical scales. To use consistent parameters $\alpha$ and $\beta$ in both modalities, we first perform a data normalization.

Rescaling the RNA coordinates by a factor $s>0$ induces a rescaling of the kinetic energy in RNA space by a factor of $s^2$, but leaves the growth cost unchanged. Specifically, let $S_s(x)=sx$ and
$\widetilde\rho_t=(S_s)_{\#}\rho_t$. The rescaled velocity and growth become
$
\widetilde u_t(S_s(x))=s\,u_t(x)
$
and
$
\widetilde g_t(S_s(x))=g_t(x).
$
By the pushforward identity,
\begin{equation*}
  \int\|\widetilde u_t\|^2\,d\widetilde\rho_t
  =s^2\int\|u_t\|^2\,d\rho_t, \quad
  \int\widetilde g_t^2\,d\widetilde\rho_t
  =\int g_t^2\,d\rho_t.
\end{equation*}

Inspired by this fact, we introduce a scale $s_x$ as below
\begin{equation*}
  s_x
  =
  \left(
    \sum_{k=0}^{K-1}
    \frac{
      W^2_2(\widehat{\rho}_k, \widehat{\rho}_{k+1})
    }{t_{k+1}-t_k}
  \right)^{1/2},
  \quad
  \widehat{\rho}_k = \sum_{i=1}^{M_k} \frac{1}{M_k}\delta_{\widehat{X}_{k}^{(i)}}, 
\end{equation*}
where $W_2^2$ is the squared 2-Wasserstein distance with a squared Euclidean cost. Each snapshot is then rescaled by this factor
\begin{equation*}
\widetilde\rho_k
=
(S_{1/s_x})_\#\widehat\rho_k
=
\frac{1}{M_k}\sum_{i=1}^{M_k}
\delta_{\widehat X_k^{(i)}/s_x}.
\end{equation*}
This rescaling gives
\begin{equation*}
  \sum_{k=0}^{K-1}
  \frac{
    W_2^2(\widetilde \rho_k,\widetilde \rho_{k+1})
  }{t_{k+1}-t_k}
= 1 .
\end{equation*}

An analogous scale \(s_y\) is computed from the ATAC snapshots. We divide the RNA and ATAC coordinates by \(s_x\) and \(s_y\), respectively, so that the cumulative reference transport cost defined above equals 1 in each modality. This provides a common reference transport scale while leaving the growth cost unchanged, which allows us to use common values of \(\alpha\) and \(\beta\) in the normalized formulation. For downstream analyses in the original PCA, LSI or PoissonVI spaces, the computed RNA and ATAC trajectories are transformed back by multiplying their coordinates by the rescaling factors \(s_x\) and \(s_y\), respectively.

For simplicity of notation, we continue to denote the rescaled empirical distributions by \(\widehat\rho_k\) and $\widehat \xi_k$.

\subsection*{Cross-modality map $\mathbf{T}_t^{\eta}$}
\label{sec:method-cross-modal-map}
We trained the cross-modality map $\mathbf{T}_t^{\eta}$ by using paired RNA and ATAC representations from multiome cells or computationally paired metacells. The map was trained before trajectory learning, and its parameters were held fixed throughout trajectory optimization.

We parameterized $\mathbf{T}_t^{\eta}$ using Feature-wise Linear Modulation (FiLM)\cite{perez2018film}, then minimized
\begin{equation*}
\mathcal L_{\mathrm{MSE}}
=
\sum_{k=1}^{K}\sum_{i=1}^{M_k}
\left\|\mathbf T_{t_k}^{\eta}(\widehat X_{k}^{(i)})- \widehat Y_k^{(i)}\right\|_2^2.
\end{equation*}
Optimization was performed for 300 epochs using AdamW, with a learning rate of $10^{-3}$, a weight decay of $10^{-4}$, and a mini-batch size of 1024.

We compared the trained FiLM map with other methods including Gaussian-kernel maps, ridge regression, and multilayer perceptrons (MLPs) with and without time input.
Evaluation includes the mean squared error between predicted ATAC coordinates and their paired observed ATAC coordinates, the Sinkhorn divergence between the mapped and observed ATAC distributions at each stage, and the sensitivity of predicted ATAC coordinates to small perturbations of the RNA inputs (Supplementary Note 6).

FiLM achieved competitive fits to the observed ATAC distributions across datasets. In the trajectory benchmark scMultiSim, synchronization also improved performance like reducing lineage error when alternative mapping architectures were used for $\mathbf{T}_t^{\eta}$, so the choice of map is not restricted to FiLM and can be adapted to the dataset.

\subsection*{\texorpdfstring{Leave-one-stage-out training and evaluation}{Leave-one-stage-out training and evaluation}}
\label{sec:method-loso-protocol}

In each leave-one-stage-out experiment, we withheld one observation stage from both cross-modality map training $\mathbf{T}_t^{\eta}$ and trajectory learning, then predicted cell states at the withheld time point. For COATI and the RNA-only model, predicted RNA states were evaluated directly, and the corresponding ATAC states were obtained using the trained cross-modality map \(\mathbf T_t^\eta\). scMultiNODE predictions were decoded into RNA and ATAC spaces using its own decoders. For each modality, all methods were evaluated in the same coordinate space using the three metrics defined below.

\subsubsection*{Distribution-level evaluation metrics}
\label{sec:method-distribution-metrics}

At a held-out time \(t_k\), let \(\{X_{t_k}^{(i)},w_{t_k}^{(i)}\}_{i=1}^{B}\) denote the predicted RNA cell states and their mass weights. Before evaluation, we normalize the weights:
\begin{align}\label{eqn:weight_normalization}
\bar w_{t_k}^{(i)}
=
\frac{w_{t_k}^{(i)}}{\sum_{l=1}^{B}w_{t_k}^{(l)}}.
\end{align}
The normalized predicted and observed RNA distributions are
\[
\overline\rho_{t_k}
=
\sum_{i=1}^{B}\bar w_{t_k}^{(i)}
\delta_{X_{t_k}^{(i)}},
\qquad
\widehat\rho_k
=
\frac{1}{M_k}
\sum_{j=1}^{M_k}
\delta_{\widehat X_k^{(j)}}.
\]
Balanced methods use uniform weights, \(\bar w_{t_k}^{(i)}=1/B\). For unbalanced methods, the predicted mass weights are normalized as above. Similarly the corresponding ATAC distributions $\overline{\xi}_{t_k}$ and $\widehat{\xi}_k$ are defined using \(Y_{t_k}^{(i)} = \mathbf{T}_{t_k}^{\eta}(X_{t_k}^{(i)})\) and \(\widehat Y_k^{(j)}\),
\[
\overline\xi_{t_k}
=
\sum_{i=1}^{B}\bar w_{t_k}^{(i)}
\delta_{Y_{t_k}^{(i)}},
\qquad
\widehat\xi_k
=
\frac{1}{N_k}
\sum_{j=1}^{N_k}
\delta_{\widehat Y_k^{(j)}}.
\]

\medskip\noindent\textbf{Local support.}
We assessed local support by measuring the predicted probability mass near each observed cell. For each observed RNA cell \(\widehat X_k^{(j)}\), we defined \(r_{k,j}^{\mathcal X}\) as its Euclidean distance to its fifteenth nearest observed RNA neighbor, excluding the cell itself. These radii define smaller neighborhoods in densely sampled regions and larger neighborhoods in sparsely sampled regions. The predicted mass within each neighborhood is
\[
a_{k,j}^{\mathcal X}
=
\sum_{i=1}^{B}\bar w_{t_k}^{(i)}
\mathbf{1}_{\left\{
\left\|X_{t_k}^{(i)}-\widehat X_k^{(j)}\right\|_2
\le r_{k,j}^{\mathcal X}
\right\}}.
\]
We converted this local mass into a support score using
\[
B_{\mathrm{eff}}(t_k)
=
\left(\sum_{i=1}^{B}
\bigl(\bar w_{t_k}^{(i)}\bigr)^2\right)^{-1},
\qquad
p_{k,j}^{\mathcal X}
=
1-\left(1-a_{k,j}^{\mathcal X}\right)^{B_{\mathrm{eff}}(t_k)}.
\]
This formulation of the support score is motivated by the probability that at least one independent draw enters the neighborhood. For fixed effective sample size $B_{\mathrm{eff}}$, the score increases with the predicted mass in the neighborhood. An empty neighborhood receives a score of zero.
For balanced methods, \(B_{\mathrm{eff}}(t_k)=B\). If \(l_{k,j}^{\mathcal X}\) predicted cells lie within the neighbourhood, then
$
p_{k,j}^{\mathcal X}
=
1-\left(1-\frac{l_{k,j}^{\mathcal X}}{B}\right)^B
$.
The overall RNA support score is
\[
\mathcal R_k^{\mathcal X}
=
\frac{1}{M_k}
\sum_{j=1}^{M_k}
p_{k,j}^{\mathcal X}.
\]
The ATAC score \(\mathcal R_k^{\mathcal Y}\) is computed analogously. Both scores lie between zero and one and summarize local coverage of the observed cell population, with higher values indicating stronger local support across the observed population.

\medskip\noindent\textbf{Distributional discrepancy.}
We computed the distributional discrepancies using the Sinkhorn divergence:
\[
D_k^{\mathcal X}
=
\sqrt{2S_{\epsilon_k^{\mathcal X}}
\!\left(\overline\rho_{t_k},\widehat\rho_k\right)},
\qquad
D_k^{\mathcal Y}
=
\sqrt{2S_{\epsilon_k^{\mathcal Y}}
\!\left(\overline\xi_{t_k},\widehat\xi_k\right)}.
\]
In each modality, the ground cost was the half squared Euclidean distance between cell coordinates. The regularization scale $\epsilon_k^{\mathcal{X},\mathcal Y}$ was determined from the median pairwise distance among observed cells, following the procedure described in \hyperref[sec:method-sinkhorn-scale-preprocessing]{Sinkhorn scale preprocessing}. Lower values indicate closer agreement between the predicted and observed distributions.
To assess the predictions at held-out stages, we also reported $\tau_k^b$ for each dataset in the comparison. This value determined the best discrepancy error can be achieved in the implementation.


\medskip\noindent\textbf{Cell-type composition.}
At each held-out time \(t_k\), each predicted cell was assigned the cell-type label of its nearest observed cell in the corresponding modality. Thus, predicted RNA states \(X_{t_k}^{(i)}\) were labelled using observed RNA cells \(\widehat X_k^{(j)}\), and predicted ATAC states \(Y_{t_k}^{(i)}\) were labelled using observed ATAC cells \(\widehat Y_k^{(j)}\). All observed reference cells had cell-type annotations.

For modality \(b\in\{\mathcal X,\mathcal Y\}\), let \(l_{k,i}^{b}\) and \(\hat l_{k,j}^{b}\) denote the assigned predicted labels and observed annotations, respectively. For each cell type \(c\), we calculated the predicted and the observed cell-type proportions at time snapshot $t_k$ as follows
\begin{align}\label{eqn:cell_type}
p_{k,c}^{b}
=
\sum_{i=1}^{B}
\bar w_{t_k}^{(i)}
\mathbf 1_{\{l_{k,i}^{b}=c\}},
\qquad
\hat p_{k,c}^{b}
=
\frac{1}{m_k^b}
\sum_{j=1}^{m_k^b}
\mathbf 1_{\{\hat l_{k,j}^{b}=c\}},
\end{align}
where \(m_k^b\) is the number of observed cells, with $m_k^{\mathcal{X}} = M_k$ and $m_k^{\mathcal{Y}} = N_k$.
We compared the resulting proportion vectors using the base-2 Jensen–Shannon divergence:
\[
\begin{aligned}
\operatorname{JSD}_2
\!\left(\boldsymbol p_k^{b},
        \hat{\boldsymbol p}_k^{b}\right)
=
\frac12\operatorname{KL}_2
\!\left(\boldsymbol p_k^{b}\middle\|
        \boldsymbol q_k^b\right)
+
\frac12\operatorname{KL}_2
\!\left(\hat{\boldsymbol p}_k^{b}\middle\|
        \boldsymbol q_k^b\right),
\quad
\boldsymbol q_k^b
=
\frac{\boldsymbol p_k^{b}
      +\hat{\boldsymbol p}_k^{b}}{2}.
\end{aligned}
\]
Cell-type composition similarity was then computed as
\[
C_k^b
=
1-\sqrt{
\operatorname{JSD}_2
\!\left(\boldsymbol p_k^{b},
        \hat{\boldsymbol p}_k^{b}\right)
}.
\]
This score ranges from zero to one, with higher values indicating closer agreement in cell-type proportions. A score of one indicates identical proportions, although it does not imply agreement in individual cell states or lineage assignments.

\subsubsection*{Reference-based estimation of gene expression and peak accessibility}
\label{sec:method-molecular-readout}

We estimated gene expression and peak accessibility along the predicted RNA and ATAC states, \(X_t^{(i)}\) and \(Y_t^{(i)}\), using neighboring observed cells. Neighbors were selected in the RNA or ATAC embedding space, or in a shared latent space, as appropriate for each analysis.
For modality \(b\in\{\mathcal X,\mathcal Y\}\), let \(\widehat{\mathcal{N}}_{t,i}^{b}\) denote the set of the observed reference cells selected for predicted cell \(i\) at time \(t\). Let \(\widehat F_{jf}^{\mathcal X}\) denote the preprocessed expression of gene \(f\) in observed reference RNA cell \(j\), and \(\widehat F_{jf}^{\mathcal Y}\) the preprocessed accessibility of peak \(f\) in observed reference ATAC cell \({j}\). We estimated the corresponding feature values as
\[\ F_{t,f}^{b,(i)}
=
\sum_{ j\in\widehat{\mathcal{N}}_{t,i}^{b}}
a_{t,i, j}^{b}\widehat F_{{j}f}^{b},
\qquad
a_{t,i,j}^{b}\geq0,
\qquad
\sum_{j\in\widehat{\mathcal N}_{t,i}^{b}}a_{t,i,j}^{b}=1.
\]
We used nearest reference cells of predicted cell $i$ at time $t$ to determine $\widehat{\mathcal N}_{t,i}^{b}$ and $a_{t,i,j}^{b}$. For peak–gene dynamics, we used Gaussian-distance weights proportional to $\exp[-(d/h)^2]$, normalized to sum to one, where $d$ is the distance to a reference cell and $h$ is the bandwidth. Unless otherwise specified, we used 15 nearest reference cells and set $h$ to be the median distance to the 15th nearest reference cell across the queried cells. Uniform weights are applied to other molecular readouts.  Thus, \(F_{t,f}^{\mathcal X,(i)}\) estimates gene expression at \(X_t^{(i)}\), and \( F_{t,f}^{\mathcal Y,(i)}\) estimates peak accessibility at \(Y_t^{(i)}\). The coefficients \(a_{t,i, j}^{b}\) are weights for averaging over observed reference neighbors. The resulting estimates remain on the preprocessing scale of the reference features.

\subsection*{Mass decomposition for unbalanced models}
\label{sec:method-mass-decomposition}

We decomposed changes in predicted cell-type proportions into three contributions: cells entering a cell type, cells leaving it and relative reweighting among cells retaining the same cell type. We tracked the same predicted cell trajectories between adjacent observation times $[t_k, t_{k+1}]$ and used their assigned RNA cell-type labels at the two endpoints $t_k$ and $t_{k+1}$.

Let \(l_{k,i}^{\mathcal X}\) denote the cell-type label assigned to \(X_{t_k}^{(i)}\), consistent with the notation in (\ref{eqn:cell_type}). We normalized the predicted mass weights over all predicted cells as in (\ref{eqn:weight_normalization}), and defined the predicted cell-type proportion $p_{k,c}^{\mathcal X}$ for cell type $c$ at time point $t_k$ as in (\ref{eqn:cell_type}).

The three contributions over \([t_k,t_{k+1}]\) were
\begin{align*}
I_{k,c}=\sum_{\substack{i:l_{k,i}^{\mathcal X}\neq c, l_{k+1,i}^{\mathcal X}=c}}\bar w_{t_{k+1}}^{(i)}, \quad
O_{k,c}=\sum_{\substack{i:l_{k,i}^{\mathcal X}=c , l_{k+1,i}^{\mathcal X}\neq c}}\bar w_{t_k}^{(i)}, \quad
R_{k,c}=\sum_{\substack{i:l_{k,i}^{\mathcal X}=c, l_{k+1,i}^{\mathcal X}=c}}\left(\bar w_{t_{k+1}}^{(i)}-\bar w_{t_k}^{(i)}\right).
\end{align*}
The entry contribution \(I_{k,c}\) uses the later-stage normalized weights, whereas the exit contribution \(O_{k,c}\) uses the earlier-stage normalized weights. The reweighting contribution \(R_{k,c}\) measures changes in the normalized weights of trajectories assigned to \(c\) at both endpoints. Partitioning the cell-type proportions according to these three groups gives the exact identity
\[
p_{k+1,c}^{\mathcal X}
-
p_{k,c}^{\mathcal X}
=
I_{k,c}-O_{k,c}+R_{k,c}.
\]
We compared this net predicted change with the observed change
$
\hat{p}_{k+1,c}^{\mathcal X}
-
\hat{p}_{k,c}^{\mathcal X},
$
calculated from the observed RNA cell-type annotations.

\subsection*{Lineage error in the simulated benchmark}
\label{sec:method-simulated-lineage-metrics}

The scMultiSim lineage tree contains two initial cell types, $\mathcal A$ and $\mathcal B$, with observations at $t_0, t_1, t_2, t_3$. Type $\mathcal A$ remains $\mathcal A$, whereas type $\mathcal B$ can differentiate to types $\mathcal C$ and $\mathcal D$. 
We discretized the trajectory interval into $M$ equal steps
and evaluated lineage consistency on the grid,
\[
\mathcal G
=
\{s_q=t_0+q\Delta t:q=1,\ldots,M\},
\qquad
\Delta t=\frac{t_3-t_0}{M}.
\]
In experiments, we set $M=60$.

We tracked each predicted cell trajectory across all observed time point. Let \(\hat{l}^{\mathcal{X}}_{0,i}\in\{\mathcal A, \mathcal B\}\) denote the initial cell type of trajectory \(i\), and let \(l_{s_q,i}^{\mathcal X}\) denote its predicted RNA cell type at \(s_q\). Let $\widehat{\mathcal N}_{s_q,i}$ be the set of the 15 nearest reference cells among all observed RNA data $\bigcup_{k=1}^K \left(\{\widehat X_{t_k}^{(i)}\}_{i=1}^{M_k}\right)$ to the predicted state $X_{s_q}^{(i)}$ at time $s_q$, and let $\widehat l_{t_k(j),j}^{\mathcal X}$ denote the annotated cell type of the $j$-th nearest reference cell, where $t_k(j)$ is the corresponding observation time for the $j$-th nearest reference cell. 
The predicted label  \(l_{s_q,i}^{\mathcal X}\) is computed as follows,
\[
l_{s_q,i}^{\mathcal X}
=
\operatorname*{arg\,max}_{c\in\{\mathcal A,\mathcal B,\mathcal C,\mathcal D\}}
\sum_{j\in\widehat{\mathcal N}_{s_q,i}}
a^{\mathcal X}_{s_q,i,j}\,
\mathbf 1_{\{\widehat l_{t_k(j),j}^{\mathcal X}=c\}},
\qquad
a^{\mathcal X}_{s_q,i,j}
\propto
\frac{1}{
\left\|X_{s_q}^{(i)}-\widehat X_{t_k(j)}^{(j)}\right\|_2
},
\]
where the neighbour weights sum to one.

We defined the lineage-compatible cell types as
\[
\mathcal C(
\mathcal A)=\{\mathcal A\},
\qquad
\mathcal C( \mathcal B)=\{\mathcal B, \mathcal C, \mathcal D\}.
\]
We reported the lineage error at the snapshot and for the whole trajectory.
The lineage error at snapshot \(t_k\) was
\[
E_{\mathrm{lineage}}(t_k)
=
\sum_{i=1}^{B}
\bar w_{t_k}^{(i)}
\mathbf 1_{\left\{
l_{t_k,i}^{\mathcal X}
\notin\mathcal C(\hat{l}^{\mathcal{X}}_{0,i})
\right\}}.
\]
As usual, balanced methods used uniform weights, whereas unbalanced methods used their normalized predicted cell masses. This error measures the proportion of predicted mass assigned to a cell type incompatible with its initial lineage.

We defined the global lineage error as
\[
E_{\mathrm{lineage}}^{\mathrm{global}}
=
\sum_{i=1}^{B}
\bar w_{t_3}^{(i)}
\mathbf 1_{\left\{
\exists s\in\mathcal G:
l_{s,i}^{\mathcal{X}}\notin\mathcal C(\hat{l}_{0,i}^{\mathcal{X}})
\right\}}.
\]
This quantity measures the fraction of terminal predicted mass associated with trajectories assigned to the other source lineage at least once on the evaluation time grid. Trajectories that subsequently returned to their source lineage were still counted as errors. For uniform weights, the score reduces to the fraction of trajectories flagged as incorrect.

\subsection*{Implementation of competing trajectory inference methods}
\label{sec:method-baselines}

We benchmarked COATI against CytoBridge\cite{zhang2025deciphering}, TrajectoryNet\cite{tong2020trajectorynet}, MIOFlow\cite{huguet2022manifold}, TIGON\cite{sha2024reconstructing} and scMultiNODE\cite{2024scmultinode}, using their public implementations. Models that required a learned representation were trained in that representation. All distributional predictions were evaluated in the same normalized RNA and ATAC coordinate spaces using the same metrics.

CytoBridge and TrajectoryNet were trained directly in the normalized RNA space used by COATI. For CytoBridge, we used the velocity-only configuration for balanced comparisons and the velocity-and-growth configuration for unbalanced comparisons. For TrajectoryNet, we compared two prediction procedures: propagating the same observed initial cells used by COATI, and generating predictions from the model's Gaussian base distribution \cite{tong2020trajectorynet}. The latter gave better distributional reconstruction in our comparisons and was therefore used for the reported distributional evaluations.

For MIOFlow, we retained its geometry-preserving representation-learning step. We trained a 10-dimensional GAGA representation from standardized RNA PCA coordinates and learned the dynamics in this representation. For TIGON, density estimation and trajectory learning directly in the higher-dimensional RNA space were computationally impractical in our experiments. We therefore trained TIGON in a 10-dimensional autoencoder representation, and kept its density-estimation and scaling procedures. Predictions from MIOFlow and TIGON were decoded and transformed back into the common normalized RNA space before distributional evaluation.

scMultiNODE learned dynamics in a shared RNA–ATAC latent space. Its predictions were converted to the common normalized RNA and ATAC spaces using its own RNA and ATAC decoders. Unless otherwise stated, biological readouts for scMultiNODE were estimated from neighboring observed reference cells in its shared latent space, following the reference-based procedures specified for each analysis.

TrajectoryNet and MIOFlow assigned consecutive internal times to the retained training stages. In each leave-one-stage-out experiment, we reassigned these internal times to consecutive internal times after removing the evaluated stage. A withheld stage was evaluated at the internal time obtained by linear interpolation between its neighbouring training stages, using their biological sampling times. Specifically, if the neighbouring stages occurred at \(t_a<t_k<t_b\) and were assigned consecutive internal times \(s_a\) and \(s_a+1\), the withheld stage \(t_k\) was evaluated at
\[
s(t_k)=s_a+\frac{t_k-t_a}{t_b-t_a}.
\]
The inverse piecewise-linear mapping was used to express the inferred trajectories in biological time.

\section*{Code availability}
COATI is available as an open-source Python package at \url{https://github.com/shankswang953/COATI}.  The code for reproducing the results with external models is available at \url{https://github.com/shankswang953/TraInfBench}.

\section*{Data availability}
\label{sec:data-availability}
The processed biological datasets assembled for this study are available on Zenodo at \url{https://zenodo.org/records/22865714}.
The synthetic datasets are available in the COATI GitHub repository at \url{https://github.com/shankswang953/COATI}.
The original sources of the biological datasets are listed below.

The mouse gastrulation RNA, peak and TSS count objects were obtained through the Bioconductor \texttt{MouseGastrulationData}/ExperimentHub
interface using \texttt{RAMultiomeData}
\cite{griffiths2026mousegastrulationdata}, and the original multiome data are available in GEO under accession \href{https://www.ncbi.nlm.nih.gov/geo/query/acc.cgi?acc=GSE205117}{GSE205117}.
The mouse secondary palate multiome data \cite{yan2024single} are available in GEO under accession
\href{https://www.ncbi.nlm.nih.gov/geo/query/acc.cgi?acc=GSE218576}{GSE218576}.
The processed mouse pancreatic endocrinogenesis multiome data
\cite{klein2025mapping} were obtained through
\texttt{moscot.datasets.pancreas\_multiome}, and the original data
are available in GEO under accession
\href{https://www.ncbi.nlm.nih.gov/geo/query/acc.cgi?acc=GSE275562}{GSE275562}.
The authors' processed Seurat objects for the human cerebral organoid
time-course data \cite{fleck2023inferring} are available on Zenodo at
\href{https://doi.org/10.5281/zenodo.5242913}{10.5281/zenodo.5242913},
and the original scRNA-seq and scATAC-seq data are available in
ArrayExpress under accessions
\href{https://www.ebi.ac.uk/biostudies/arrayexpress/studies/E-MTAB-12001}{E-MTAB-12001}
and
\href{https://www.ebi.ac.uk/biostudies/arrayexpress/studies/E-MTAB-11998}{E-MTAB-11998},
respectively.

\nolinenumbers

\bibliography{sn-bibliography}

\section*{AI Declaration}
AI tools were used to assist with manuscript editing and code organization. The authors have fully reviewed AI-assisted contents.

\section*{Acknowledgement}
This work was partly supported by NSF grant DMS2142500 (Y.Z.), NSF grant DMS2151934 (Z.C.) and
NIH grant R01GM152494 (Q.N. and Z.C.).

\section*{Competing Interests}
The authors declare no competing interests.

\end{document}